\documentclass[12pt,draftcls,journal,onecolumn]{IEEEtran}

\usepackage{graphicx}
\usepackage[cmex10]{amsmath}
\usepackage{amsfonts}
\usepackage{amssymb}
\usepackage{mathrsfs}
\usepackage{bm}
\usepackage{algorithmic}
\usepackage{algorithm}
\usepackage{array}
\usepackage{mdwmath}
\usepackage{mdwtab}
\usepackage[caption=false,font=footnotesize]{subfig}
\usepackage{fixltx2e}
\usepackage{stfloats}
\usepackage{footnote}
\usepackage{tabularx}
\usepackage{multirow}
\usepackage{color}
\usepackage{textcomp}

\newtheorem{definition}{Definition}
\newtheorem{proposition}{Proposition}
\newtheorem{theorem}{Theorem}
\newtheorem{lemma}{Lemma}

\newtheorem{remark}{Remark}

\begin{document}

\title{Dynamic Spectrum Sharing Among Repeatedly Interacting Selfish Users With Imperfect Monitoring}

\author{Yuanzhang~Xiao and~Mihaela~van~der~Schaar,~\IEEEmembership{Fellow,~IEEE}
\thanks{Manuscript received January 5, 2012; revised May 16, 2012; accepted July 2, 2012.}
\thanks{Y.~Xiao and M.~van der Schaar are with Department of Electrical Engineering, UCLA. Email: \{yxiao,mihaela\}@ee.ucla.edu.}
}

\maketitle

\begin{abstract}
We develop a novel design framework for dynamic distributed spectrum sharing among secondary users (SUs), who adjust their power levels to compete
for spectrum opportunities while satisfying the interference temperature (IT) constraints imposed by primary users. The considered interaction among
the SUs is characterized by the following three unique features. First, the SUs are interacting with each other repeatedly and they can coexist in
the system for a long time. Second, the SUs have limited and imperfect monitoring ability: they only observe whether the IT constraints are violated,
and their observation is imperfect due to the erroneous measurements. Third, since the SUs are decentralized, they are selfish and aim to maximize
their own long-term payoffs from utilizing the network rather than obeying the prescribed allocation of a centralized controller. To capture these
unique features, we model the interaction of the SUs as a repeated game with imperfect monitoring. We first characterize the set of Pareto optimal
operating points that can be achieved by deviation-proof spectrum sharing policies, which are policies that the selfish users find it in their
interest to comply with. Next, for any given operating point in this set, we show how to construct a deviation-proof policy to achieve it. The
constructed deviation-proof policy is amenable to distributed implementation, and allows users to transmit in a time-division multiple-access (TDMA)
fashion. In the presence of strong multi-user interference, our policy outperforms existing spectrum sharing policies that dictate users to transmit
at constant power levels simultaneously. Moreover, our policy can achieve Pareto optimality even when the SUs have limited and imperfect monitoring
ability, as opposed to existing solutions based on repeated game models, which require perfect monitoring abilities. Simulation results validate our
analytical results and quantify the performance gains enabled by the proposed spectrum sharing policies.
\end{abstract}


\section{Introduction}
Cognitive radios have increased in popularity in recent years, because they have the potential to significantly improve the spectrum efficiency.
Specifically, cognitive radios enable the secondary users (SUs), who initially have no rights to use the spectrum, to share the spectrum with primary
users (PUs), who are licensed to use the spectrum, as long as the PUs' quality of service (QoS), such as the throughput, is not affected by the SUs
\cite{Haykin}. A common approach to guarantee PUs' QoS requirements is to impose \emph{interference temperature} (IT) constraints
\cite{Haykin}\cite{KangZhangLiang}\cite{TanLow}\cite{XingChandramouli}--\cite{SharmaTeneketzis_GameTheory}; that is, the SUs cannot generate an
interference level higher than the interference temperature limit set by the PUs. One of the major challenges in designing cognitive radio systems is
to construct a spectrum sharing policy that achieves high spectrum efficiency while maintaining the IT constraints set by PUs.

The spectrum sharing policy, which specifies the SUs' transmit power levels, is essential to improve spectrum efficiency and protect the PUs' QoS.
Since SUs can use the spectrum as long as they do not degrade the PUs' QoS, they can use the spectrum and coexist in the system for long periods of
time. In general, the optimal spectrum sharing policy should allow SUs to transmit at different power levels temporally even when the environment
(e.g. the number of SUs, the channel gains) remains unchanged. However, most existing spectrum sharing policies require the SUs to transmit at
\emph{constant} power levels over the time horizon in which they interact\footnote{Although some spectrum sharing policies go through a transient
period of adjusting the power levels before the convergence to the optimal power levels, the users maintain constant power levels after the
convergence.} \cite{KangZhangLiang}--\cite{XiaovanderSchaar_PowerControl}. These policies with constant power levels are inefficient in many spectrum
sharing scenarios where the interference among the SUs is strong. Under strong multi-user interference, increasing one user's power level
significantly degrades the other users' QoS. Hence, when the cross channel gains are large, the feasible QoS region is nonconvex
\cite{StanczakBoche_IT07}. In this case of nonconvex feasible QoS region, a spectrum sharing policy with constant power levels is inferior to a
policy with \emph{time-varying} power levels in which the users transmit in a time-division multiple-access (TDMA) fashion, because the latter can
achieve the Pareto boundary of the convex hull of the nonconvex feasible QoS region.

Another important feature neglected in the design of spectrum sharing policies in recent works \cite{KangZhangLiang}--\cite{SorooshyariTanChiang} is
the selfishness of SUs, who aim to maximize their own QoS and may deviate from the prescribed spectrum sharing policy, if by doing so their QoS can
be improved. Hence, the spectrum sharing policy should be \emph{deviation-proof}, which means that selfish SUs cannot improve their QoS by deviating
from the policy. In this way, selfish SUs will find it in their self-interest to follow the policy.

Given the fact that the SUs will interact with each other repeatedly when sharing the spectrum, we model the interaction among the SUs as a repeated
game. In a repeated game, the stage game is played repeatedly, and a user's payoff in the repeated game is the discounted average of the stage-game
payoffs (i.e. QoS in the stage games). Users can choose different actions (i.e. power levels) in different stage games, and the repeated-game payoff
is a convex combination of different stage-game payoffs. A repeated-game strategy prescribes what action to take given past observations, and
therefore, can be considered as a spectrum sharing policy. If a repeated game strategy constitutes an equilibrium, then no user can gain from
deviation at any occasion. Hence, an equilibrium strategy is a deviation-proof spectrum sharing policy.

The spectrum sharing policy in a repeated game framework was studied in \cite{EtkinTse}--\cite{XiaoMihaela_RepeatedGame}, under the assumption of
\emph{perfect} monitoring, namely the assumption that each SU can perfectly monitor the individual transmit power levels of all the other SUs. In the
policies in \cite{EtkinTse}--\cite{XiaoMihaela_RepeatedGame}, when a deviation from the prescribed policy by any user is detected, a perpetual
punishment phase \cite{EtkinTse} or a punishment phase of certain duration \cite{WuWangLiu}\cite{XiaoMihaela_RepeatedGame} will be triggered. In the
punishment phase, all the users transmit at the maximum power levels to create strong interference to each other, resulting in low QoS of all the
users as a punishment. Due to the threat of this punishment, all the users will follow the policy in their self-interests. However, since the
monitoring can never be perfect, the punishment phase, in which all the users receive low throughput, will be triggered even if no one deviates.
Thus, the users' repeated-game payoffs, averaged over all the stage-game payoffs, cannot be Pareto optimal because of the low payoffs received in the
punishment phases. Hence, the policies in \cite{EtkinTse}--\cite{XiaoMihaela_RepeatedGame} must have performance loss in practice where the
monitoring is always imperfect.

Repeated games with imperfect monitoring have been studied extensively in the game theory literature. In \cite{FudenbergLevineMaskin94}, it is shown
that for a general repeated game with imperfect monitoring, Pareto optimal operating points can be asymptotically achieved if certain sufficient
conditions are satisfied. One sufficient condition requires the users to be able to statistically distinguish sufficiently many different actions.
Translated to the spectrum sharing scenario, it requires the SUs to be able to distinguish a certain number of interference temperature levels, where
the number of distinguishable IT levels grows linearly with the number of power levels each user can choose from. This requirement indicates the need
for a large amount of feedback information on IT levels. Moreover, another sufficient condition requires the users to be sufficiently patient, namely
they discount future payoffs arbitrarily little (i.e., their discount factors are arbitrarily close to one). This requirement on the users' patience
limits the scenarios to which the policy in \cite{FudenbergLevineMaskin94} can be applied.

In this paper, we design deviation-proof spectrum sharing policies with time-varying power levels to achieve Pareto optimal operating points that are
not achievable by existing policies with constant power levels \cite{KangZhangLiang}--\cite{XiaovanderSchaar_PowerControl}. We provide a systematic
design approach, which first characterizes the set of Pareto optimal operating points achievable by deviation-proof policies, and then for any
operating point in this set, constructs a deviation-proof policy to achieve it. The proposed policy can be easily implemented in a distributed
manner. Moreover, we prove that the proposed policy can achieve Pareto optimal operating points, even when the SUs are \emph{impatient} (namely they
discount future payoffs, and their discount factor are strictly smaller than one), and have \emph{limited} and \emph{imperfect} monitoring ability.
Specifically, their monitoring ability can be limited in that they only need to distinguish two IT levels regardless of the number of power levels
each user can choose from, and their monitoring can be imperfect due to the erroneous measurements of the interference temperature.\footnote{As will
be described later in this paper, there is an entity that regulates the interference temperature in the system, who measures the interference
temperature imperfectly and feedbacks to the users a binary signal indicating whether the constraints are violated.} This requirement on the users'
monitoring ability is significantly relaxed compared to existing works based on repeated games, which require either perfect monitoring of all the
users' individual transmit power levels \cite{EtkinTse}--\cite{XiaoMihaela_RepeatedGame} or sufficiently good monitoring to distinguish sufficiently
many IT levels \cite{FudenbergLevineMaskin94}.

\begin{figure}
\centering
\includegraphics[width =3.5in]{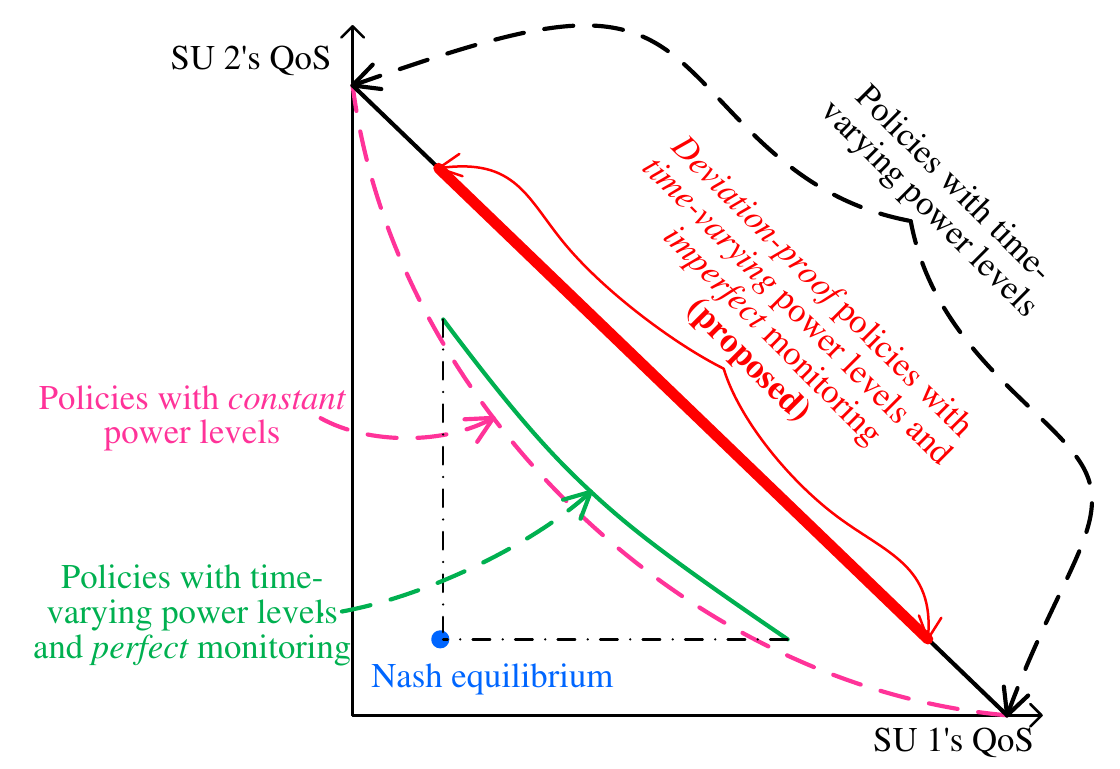}
\caption{An illustration of the best operating points achievable by different policies in a two-SU spectrum sharing system.}
\label{fig:TheBigPicture}
\end{figure}

We illustrate the performance gain of the proposed policies over the existing policies in Fig.~\ref{fig:TheBigPicture}. We show the best operating
points achievable by different classes of policies in a spectrum sharing system with two SUs. Due to the strong multi-user interference, the best
operating points achievable by policies with constant power levels \cite{KangZhangLiang}--\cite{XiaovanderSchaar_PowerControl} (the dashed curve) are
Pareto dominated by the best operating points achieved by policies with time-varying power levels (the straight line). The proposed policy, which are
deviation-proof, can achieve a portion of the Pareto optimal operating points (the thick line). Under imperfect monitoring, the policies designed
under the assumption of perfect monitoring \cite{EtkinTse}--\cite{XiaoMihaela_RepeatedGame} (the solid curve) have large performance loss compared to
the proposed policy.

\begin{table}
\renewcommand{\arraystretch}{1.1}
\caption{Comparison With Related Works In Dynamic Spectrum Sharing.} \label{table:RelatedWork} \centering
\begin{tabular}{|p{1.1cm}|c|c|p{1.2cm}|c|}
\hline
 & Power levels & Distributed & Deviation-proof & Monitoring \\
\hline
\cite{KangZhangLiang}\cite{TanLow} & Constant & No & No & N/A \\
\hline
\cite{HuangBerry_JSAC06}--\cite{SorooshyariTanChiang} & Constant & Yes & No & N/A \\
\hline
\cite{HuangBerry_06_Auction}--\cite{XiaovanderSchaar_PowerControl} & Constant & Yes & Yes & N/A \\
\hline
\cite{EtkinTse}--\cite{XiaoMihaela_RepeatedGame} & Time-varying & Yes & Yes & Imperfect \\
\hline
Proposed & Time-varying & Yes & Yes & Perfect \\
\hline
\end{tabular}
\end{table}

Finally, we summarize the comparison of our work with the existing works in dynamic spectrum sharing in Table~\ref{table:RelatedWork}. We distinguish
our work from existing works in the following categories: the power levels prescribed by the spectrum sharing policy are constant or time-varying,
whether the policy can be implemented in a distributed fashion or not, whether the policy is deviation-proof or not, and what are the requirements on
the SUs' monitoring ability. The ``monitoring'' category is only discussed within the works based on repeated games.

The rest of the paper is organized as follows.
In Section~\ref{sec:Model}, we describe the system model for dynamic spectrum sharing. Then, in Section~\ref{sec:Formulation}, we formulate the
policy design problem using repeated games. We solve the policy design problem in Section~\ref{sec:Solution}. Simulation results are presented in
Section~\ref{sec:Simulation}. Finally, Section~\ref{sec:Conclusion} concludes the paper.

%
%

\section{System Model For Dynamic Spectrum Sharing}\label{sec:Model}
\begin{figure}
\centering
\includegraphics[width =3.5in]{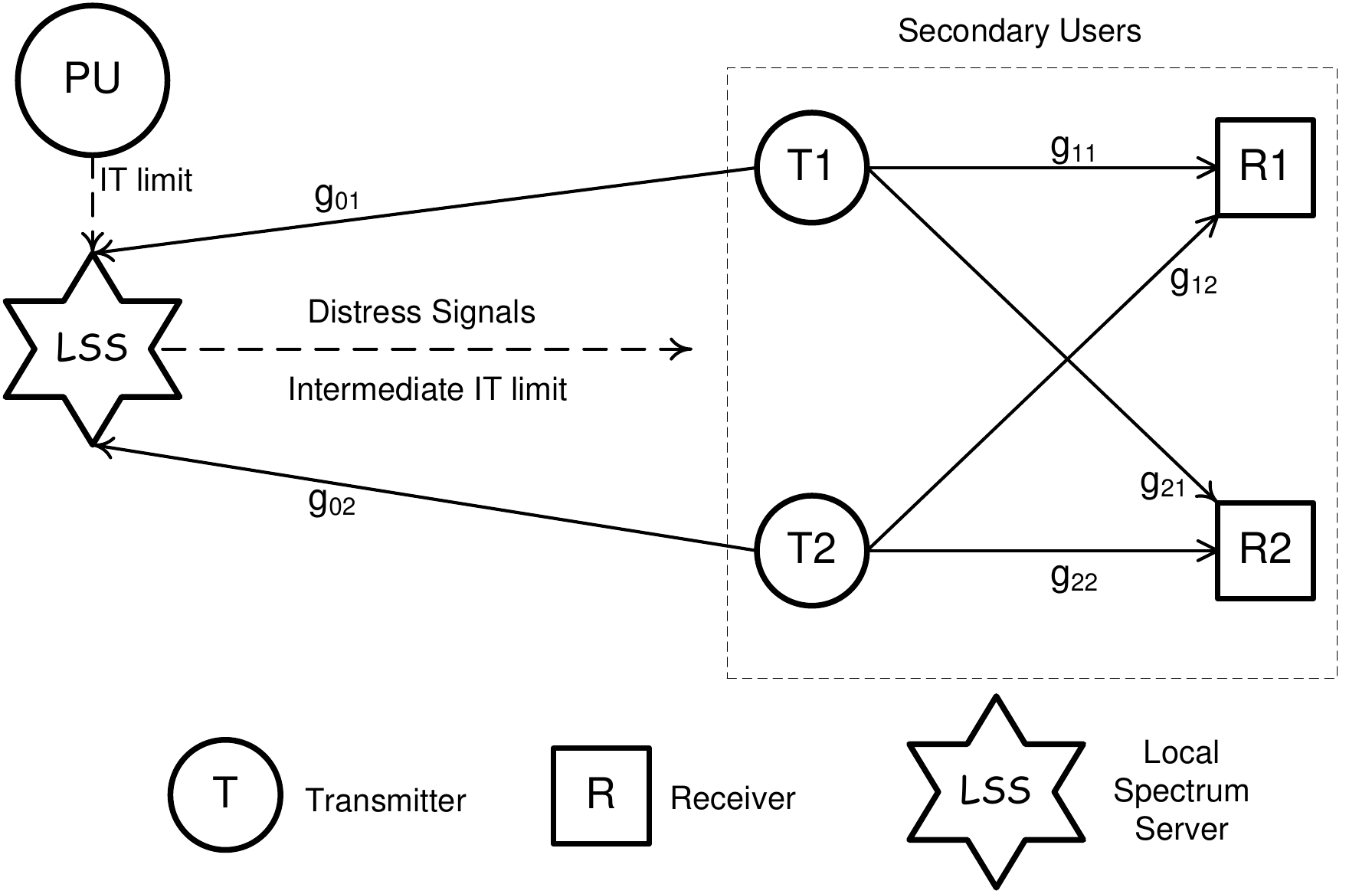}
\caption{An example system model with two secondary users. The solid line represents a link for data transmission, and the dashed line indicate a
link for control signals. The channel gains for the corresponding data link are written in the figure. The primary user (PU) specifies the
interference temperature (IT) limit to the local spectrum server (LSS). The LSS sets the intermediate IT limit to the secondary users and send
distress signals if the estimated interference power exceeds the IT limit.} \label{fig:SystemModel}
\end{figure}

We consider a system with one primary user \footnote{Although we study a system with one PU as in
\cite{KangZhangLiang}--\cite{TanLow}\cite{XingChandramouli}--\cite{GatsisMarquesGiannakis}\cite{HuangBerry_06_Auction}, our model and design
framework can be easily extended to the scenario of multiple PUs located in different geographic regions.} and $N$ secondary users (see
Fig~\ref{fig:SystemModel} for an illustrating example of a system with two secondary users). The set of SUs is denoted by $\mathcal{N} \triangleq
\{1,2,\ldots,N\}$. Each SU has a transmitter and a receiver. The channel gain from SU $i$'s transmitter to SU $j$'s receiver is $g_{ij}$. Each SU $i$
chooses a power level $p_i$ from a finite set $\mathcal{P}_i$. In other words, each SU choose from discrete power levels. We assume that
$0\in\mathcal{P}_i$, namely SU $i$ can choose not to transmit. We define SU $i$'s maximum transmit power as $P_i^{\rm max}=\max_{p_i\in\mathcal{P}_i}
p_i$. The set of joint power profiles is denoted by $\mathcal{P}=\prod_{i\in\mathcal{N}} \mathcal{P}_i$, and the joint power profile of all the SUs
is denoted by ${\bf p}=(p_1,\ldots,p_N) \in \mathcal{P}$. Let ${\bf p}_{-i}$ be the power profile of all the SUs other than SU $i$. Each SU $i$'s
instantaneous payoff (QoS) is a function of the joint power profile, namely $u_i:\mathcal{P}\rightarrow\mathbb{R}^+$. Each SU $i$'s payoff
$u_i(\mathbf{p})$ is decreasing in the other SUs' power levels $p_j,~\forall j\neq i$. Note that we do \emph{not} assume that $u_i(\mathbf{p})$ is
increasing in $p_i$.\footnote{In some scenarios with energy efficiency considerations, the payoff is defined as the ratio of throughput to transmit
power, which may not monotonically increase with the transmit power.} But we do assume that $u_i(\mathbf{p})=0$ if $p_i=0$, because a SU's payoff
should be zero when it does not transmit. One example of many possible payoff functions is the SU's throughput:
\begin{eqnarray}\label{eqn:Throughput}
u_i(\mathbf{p}) = \log_2\Big(1+\frac{p_ig_{ii}}{\sum_{j\in\mathcal{N},j\neq i} p_j g_{ji}  + n_i}\Big),
\end{eqnarray}
where $n_i$ is the noise power at SU $i$'s receiver.

As in \cite{IleriMandayam}--\cite{SorooshyariTanChiang}, there is a local spectrum server (LSS) serving as a mediating entity among the SUs. The LSS
has a receiver to measure the interference temperature and a transmitter to broadcast signals, but it cannot control the actions of the autonomous
SUs. The LSS could be a device deployed by the PU or simply the PU itself, if the PU manages by itself the spectrum leased to the SUs. Even when the
PU is the LSS, it is beneficial to consider the LSS as a separate logical entity that performs the functionality of spectrum management. The LSS
could also be a device deployed by some regulatory agency such as Federal Communications Commission (FCC), who uses it for spectrum management in
that local geographic area. In both cases, the LSS aims to improve the spectrum efficiency (e.g. the sum throughput of all the SUs) and the fairness,
while ensuring that the IT limit set by the PU is not violated. Note that the PU may also want to maximize the spectrum efficiency to maximize its
revenue obtained from spectrum leasing, since its revenue may be proportional to the sum throughput of the SUs.

The LSS measures the interference temperature at its receiver imperfectly. The measurement can be written as $\sum_{i\in\mathcal{N}} p_i g_{i0} +
\varepsilon$, where $g_{i0}$ is the channel gain from SU $i$'s transmitter to the LSS's receiver, and $\varepsilon$ is the additive measurement
error. We assume that the measurement error has zero mean and a probability distribution function $f_{\varepsilon}$ known to the LSS. We assume as in
most existing works (e.g. \cite{KangZhangLiang}--\cite{HuangBerry_06_Auction}) that the IT limit $\bar{I}$ set by the PU is known perfectly by the
LSS. Although the LSS aims to keep the interference temperature below the IT limit $\bar{I}$, it will set a lower intermediate IT limit
$I\leq\bar{I}$ to be conservative because of measurement errors. Hence, the IT constraint imposed by the LSS is
\begin{eqnarray}\label{eqn:ITC}
\begin{array}{c}\sum_{i\in\mathcal{N}} p_i g_{i0} \leq I.\end{array}
\end{eqnarray}
Even if the actual interference temperature $\sum_{i\in\mathcal{N}} p_i g_{i0}$ does not exceed the intermediate IT limit $I$, the erroneous
measurement $\sum_{i\in\mathcal{N}} p_i g_{i0} + \varepsilon$ may still exceed the IT limit $\bar{I}$ set by the PU. In this case, the LSS will
broadcast a distress signal to all the SUs. Given the joint power profile $\mathbf{p}$, this false alarm probability is
\begin{eqnarray}\label{eqn:FalseAlarmProb}
\begin{array}{c}\Gamma(\mathbf{p}) = \Pr\left(\sum_{i\in\mathcal{N}} p_i g_{i0}+\varepsilon>\bar{I}~|\sum_{i\in\mathcal{N}} p_i g_{i0} \leq
I\right),\end{array}
\end{eqnarray}
where $\Pr(A)$ is the probability that the event $A$ happens. We can see that a larger intermediate IT limit $I$ enables the SUs to transmit at
higher power levels, but results in a larger false alarm probability and a higher frequency of sending distress signals. Hence, there is an
interesting tradeoff between the spectrum efficiency and the cost of sending distress signals.

A SU's payoff is affected by the multi-user interference $\sum_{j\in\mathcal{N},j\neq i} p_j g_{ji}$, which is dependent on the cross channel gains
among different SUs. When the multi-user interference is weak due to small cross channel gains, power control becomes less important, since one SU's
power level does not affect the others' payoffs. Hence, in this paper, we focus on the more interesting scenario when the multi-user interference is
strong and power control is essential for efficient interference management. We quantify the strength of multi-user interference as follows. First,
we write $\mathbf{\tilde{p}}^i=(\tilde{p}_1^i,\ldots,\tilde{p}_N^i)$ as the joint power profile that maximizes SU $i$'s payoff subject to the IT
constraint, namely
\begin{eqnarray}
\mathbf{\tilde{p}}^i= \arg\max_{\mathbf{p}\in \mathcal{P}} u_i(\mathbf{p}),~\mathrm{subject~to}~\begin{array}{c}\sum_{i\in\mathcal{N}} p_i g_{i0}
\leq I.\end{array}
\end{eqnarray}
Since $u_i$ is decreasing in $p_j, \forall j\neq i$, we have $\tilde{p}_j^i = 0,~\forall j\neq i.$ For notational simplicity, we define the maximum
payoff achievable by SU $i$ as $\bar{v}_i \triangleq u_i(\mathbf{\tilde{p}}^i)$. Then, we say a spectrum sharing scenario has strong multi-user
interference if the following property is satisfied.
\begin{definition}[Strong Multi-user Interference]\label{definition:strong_interference}
A spectrum sharing scenario has strong multi-user interference, if the set of feasible payoffs
$\mathcal{V}=\mathrm{conv}\{\mathbf{u}(\mathbf{p})=(u_1(\mathbf{p}),\ldots,u_N(\mathbf{p})): \mathbf{p}\in \mathcal{P},\sum_{i\in\mathcal{N}} p_i
g_{i0} \leq I\}$, where $\mathrm{conv}(X)$ is the convex hull of $X$, has $N+1$ extremal points\footnote{The extremal points of a convex set are
those that are not convex combinations of other points in the set.}: $(0,\ldots,0)\in\mathbb{R}^N$,
$\mathbf{u}(\mathbf{\tilde{p}}^1),\ldots,\mathbf{u}(\mathbf{\tilde{p}}^N)$.
\end{definition}
This definition characterizes the strong interference among the SUs: the increase of one SU's payoff comes at such an expense of the other SUs'
payoffs that the set of feasible payoffs without time sharing is nonconvex. A spectrum sharing scenario satisfies this property when the cross
channel gains among users are large \cite{StanczakBoche_IT07}. In the extreme case of strong multi-user interference, simultaneous transmissions from
different SUs result in packet loss, as captured in the collision model \cite{ParkvanderSchaar_CognitiveMAC}. According to this definition, the set
of feasible payoffs can be written as $\mathcal{V}=\mathrm{conv}\{(0,\ldots,0),
\mathbf{u}(\mathbf{\tilde{p}}^1),\ldots,\mathbf{u}(\mathbf{\tilde{p}}^N)\}$. Moreover, its Pareto boundary is
$\mathcal{B}=\{\mathbf{v}\in\mathcal{V}:\sum_{i=1}^N v_i/\bar{v}_i=1,~v_i\geq0, \forall i\}$ as part of a hyperplane, which can be achieved only by
SUs transmitting in a TDMA fashion.

\section{Formulation of The Policy Design Problem}\label{sec:Formulation}
In this section, we first formulate the interaction among the SUs as a repeated game with imperfect monitoring, and define the deviation-proof
spectrum sharing policy. Then, we formally define the policy design problem and outline our design framework to solve it.

\subsection{Formulation of The Repeated Game}
Similar to \cite{KangZhangLiang}--\cite{XiaovanderSchaar_PowerControl}, we assume that the system parameters, such as the number of SUs and the
channel gains, remain fixed during the considered time horizon. The system is time slotted at $t=0,1,\ldots$. We assume that the users are
synchronized as in \cite{KangZhangLiang}--\cite{XiaovanderSchaar_PowerControl}. At the beginning of time slot $t$, each SU $i$ chooses its power
level $p_i^t$, and receives a payoff $u_i(\mathbf{p}^t)$. The LSS obtains the measurement $\sum_{i\in\mathcal{N}} p_i^t g_{i0}+\varepsilon^t$, where
$\varepsilon^t$ is the realization of the error $\varepsilon$ at time slot $t$, and compare the measurement with the IT limit $\bar{I}$. The set of
measurement outcomes of the comparison $Y$ has two elements, namely $Y=\{y_0,y_1\}$. The (measurement) outcome $y^t$ is determined by
\begin{eqnarray}
y^t=\left\{\begin{array}{ll}y_0,&\mathrm{if}~\sum_{i\in\mathcal{N}} p_i^t g_{i0}+\varepsilon^t > \bar{I} \\
y_1,&\mathrm{otherwise}\end{array}\right..
\end{eqnarray}
We write the conditional probability distribution of the outcome $y$ given the joint power profile $\mathbf{p}$ as $\rho(y|\mathbf{p})$, which can be
calculated as
\begin{eqnarray}\label{eqn:ConditionalDistribution_PowerControl}
\rho(y_1|\mathbf{p}) &=& \int_{x\leq \bar{I}-\sum_{i\in\mathcal{N}} p_i g_{i0}} f_{\varepsilon}(x)~dx, \nonumber \\
\rho(y_0|\mathbf{p}) &=& 1-\rho(y_1|\mathbf{p}).
\end{eqnarray}
At the end of time slot $t$, the LSS sends a distress signal if the outcome $y^t=y_0$. Note that the LSS does not send signals when the outcome is
$y_1$, and the SUs know that the outcome is $y_1$ by default when they do not receive the distress signal.

Note that in repeated games with perfect monitoring \cite{EtkinTse}--\cite{XiaoMihaela_RepeatedGame}, the outcome available to each SU at time slot
$t$ is precisely the joint power profile chosen by the SUs, i.e. $y^t=\mathbf{p}^t$. We say the monitoring is imperfect if $y^t\neq\mathbf{p}^t$. In
a general repeated game with imperfect monitoring, in order to achieve Pareto optimality, the set of outcomes $Y$ should have a large cardinality,
namely $|Y|\geq |\mathcal{P}_i|+|\mathcal{P}_j|-1$ for all $i\in\mathcal{N}$ and all $j\neq i$ \cite{FudenbergLevineMaskin94}. In contrast, our
proposed policy can achieve Pareto optimality even when $|Y|=2$ regardless of the cardinality of the SU's action set $\mathcal{P}_i$.

At each time slot $t$, each SU $i$ determines its transmit power $p_i^t$ based on its history, which is a collection of all the past power levels it
has chosen and all the past measurement outcomes. Formally, the history of SU $i$ up to time slot $t\geq1$ is
$h_i^t=\{p_i^0,y^0;\ldots;p_i^{t-1},y^{t-1}\}\in (\mathcal{P}_i\times Y)^t$, and that at time slot $0$ is $h_i^0=\varnothing$. The history of SU $i$
contains private information about SU $i$'s power levels that is unknown to the other SUs; in contrast, we define the \emph{public history} as
$h^t=\{y^0;\ldots;y^{t-1}\}\in Y^t$ for $t\geq1$ and $h^0=\varnothing$. The public history $h^t$ only contains the measurement outcomes that are
known to all the SUs.

In this paper, we focus on \emph{public strategies}, in which each SU's decision depends on the public history only. Hence, each SU $i$'s strategy
$\sigma_i$ is a mapping from the set of all possible public histories to its action set, namely $\sigma_i: \sqcup_{t=0}^\infty Y^t \rightarrow
\mathcal{P}_i$. Due to realization equivalence principle \cite[Lemma~7.1.2]{MailathSamuelson}, we lose nothing by only considering public strategies,
in terms of the achievable Pareto optimal operating points.

The spectrum sharing policy is the joint strategy profile of all the SUs, defined as $\bm{\sigma}=(\sigma_1,\ldots,\sigma_N)$. The SUs are selfish
and maximize their own long-term discounted payoffs. Assuming, as in \cite{EtkinTse}--\cite{FudenbergLevineMaskin94}, the same discount factor
$\delta\in[0,1)$ for all the SUs, each SU $i$'s (long-term discounted) payoff can be written as
\begin{eqnarray}\label{RepeatedGamePayoff}
U_i(\bm{\sigma}) = (1-\delta) \left[ u_i(\mathbf{p}^0) + \sum_{t=1}^\infty \delta^t \cdot \!\!\!\! \sum_{y^{t-1}\in Y}
\!\!\!\!\rho(y^{t-1}|\mathbf{p}^{t-1}) u_i(\mathbf{p}^t)\right], \nonumber
\end{eqnarray}
where $\mathbf{p}^0$ is determined by $\mathbf{p}^0=\bm{\sigma}(\varnothing)$, and $\mathbf{p}^t$ for $t\geq1$ is determined by
$\mathbf{p}^t=\bm{\sigma}(h^t)=\bm{\sigma}(h^{t-1};y^{t-1})$. The discount factor represents the ``patience'' of the SUs; a larger discount factor
indicates that a SU is more patient. The discount factor is determined by the delay sensitivity of the SUs' applications.


We define the deviation-proof policy as the perfect public equilibrium (PPE) of the game. The PPE prescribes a strategy profile $\bm{\sigma}$ from
which no SU has incentive to deviate after any given history at any time slot, and thus can be considered as a deviation-proof policy. It is normally
more strict than Nash equilibrium, because it requires that the SUs have no incentive to deviate at any given history, while Nash equilibrium only
guarantees this at the histories that possibly arise from the equilibrium strategy. We can also consider PPE in repeated games with imperfect
monitoring as the counterpart of subgame perfect equilibrium defined in repeated games with perfect monitoring \cite{MailathSamuelson}.

Before the definition of PPE, we introduce the concept of continuation strategy: SU $i$'s continuation strategy induced by any history $h^t\in Y^t$,
denoted $\sigma_i|_{h^t}$, is defined by $\sigma_i|_{h^t}(h^\tau)=\sigma_i(h^t h^\tau), \forall h^\tau \in Y^\tau$, where $h^t h^\tau$ is the
concatenation of the history $h^t$ followed by the history $h^\tau$. By convention, we denote $\bm{\sigma}|_{h^t}$ and $\bm{\sigma}_{-i}|_{h^t}$ the
continuation strategy profile induced by $h^t$ of all the SUs and that of all the SUs other than SU $i$, respectively. Then the PPE is defined as
follows \cite[Definition~7.1.2]{MailathSamuelson}
\begin{definition}[Perfect Public Equilibrium]
A strategy profile $\bm{\sigma}$ is a perfect public equilibrium if for any public history $h^t\in Y^t$, the induced continuation strategy
$\bm{\sigma}|_{h^t}$ is a Nash equilibrium of the continuation game, namely for all $i\in\mathcal{N}$,
\begin{eqnarray}
U_i(\bm{\sigma}|_{h^t}) \geq U_i(\sigma_i'|_{h^t},\bm{\sigma}_{-i}|_{h^t}),~\mathrm{for~all}~\sigma_i'.
\end{eqnarray}
\end{definition}
We define the equilibrium payoff as a vector of payoffs $\mathbf{v}=(U_1(\bm{\sigma}),\ldots,U_N(\bm{\sigma}))$ achieved at the equilibrium.

\subsection{The Policy Design Problem}
The primary user or the regulatory agency aims to maximize an objective function defined on the SUs' payoffs, $W(U_1(\bm{\sigma}), \ldots,
U_N(\bm{\sigma}))$. This definition of the objective function is general enough to include the objective functions deployed in many existing works,
such as \cite{KangZhangLiang}--\cite{GatsisMarquesGiannakis}\cite{EtkinTse}\cite{WuWangLiu}. An example of the objective function is the weighted sum
payoff $\sum_{i=1}^N w_i U_i$, where $\{w_i\}_{i=1}^N$ are the weights satisfying $w_i\in[0,1],\forall i$ and $\sum_{i=1}^N w_i=1$. The PU
(respectively, the regulatory agency) maximizes the objective function for the revenue (the spectrum efficiency), while maintaining the IT constraint
\eqref{eqn:ITC}. To reduce the cost of sending distress signals, a constraint on the false alarm probability is also imposed as
$\Gamma(\mathbf{p})\leq \bar{\Gamma}$, where $\bar{\Gamma}$ is the maximum false alarm probability allowed. At the maximum of the welfare function,
some SUs may have extremely low payoffs. To avoid this, a minimum payoff guarantee $\gamma_i\geq0$ is imposed for each SU $i$. To sum up, we can
formally define the policy design problem as follows
\begin{eqnarray}\label{eqn:PolicyDesignProblem}
&\displaystyle\max_{\bm{\sigma}}& W(U_1(\bm{\sigma}), \ldots, U_N(\bm{\sigma})) \\
&s.t.& \bm{\sigma}~\mathrm{is~public~perfect~equilibrium}, \nonumber\\
&    & \sum_{i\in\mathcal{N}} \sigma_i(h^t)\cdot g_{i0} \leq I,~\forall t,~\forall h^t\in Y^t,\nonumber\\
&    & \Gamma(\bm{\sigma}(h^t))\leq \bar{\Gamma},~\forall t,~\forall h^t\in Y^t,\nonumber\\
&    & U_i(\bm{\sigma})\geq \gamma_i,~\forall i\in\mathcal{N}. \nonumber
\end{eqnarray}

\section{Solving The Policy Design Problem}\label{sec:Solution}
\begin{figure}
\centering
\includegraphics[width =3.5in]{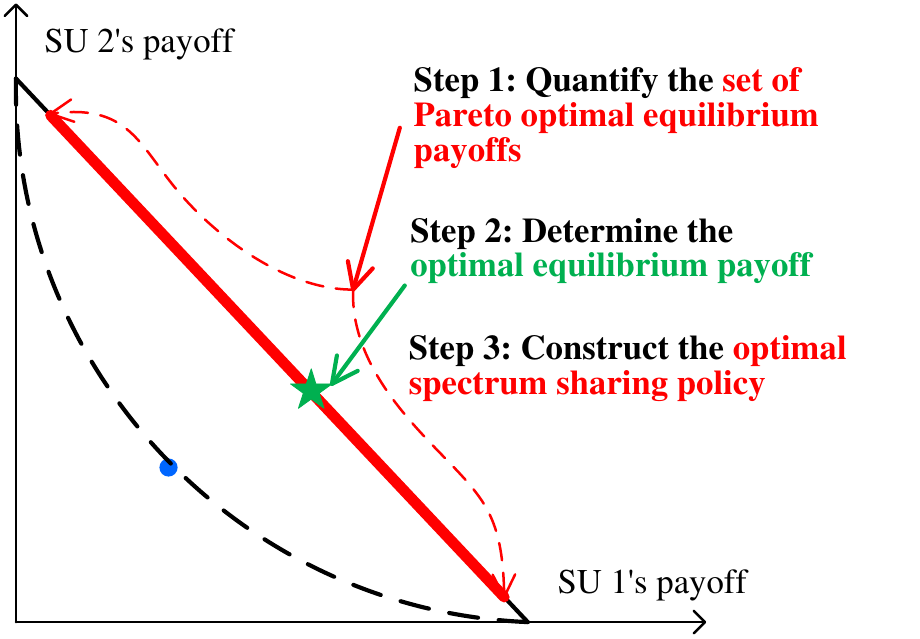}
\caption{The procedure of solving the design problem.} \label{fig:DesignFramework_StrongNegativeExternality}
\end{figure}

In this section, we solve the policy design problem \eqref{eqn:PolicyDesignProblem} following the procedure outlined in
Fig.~\ref{fig:DesignFramework_StrongNegativeExternality}. We first quantify the set of Pareto optimal equilibrium payoffs (i.e. the Pareto optimal
payoffs that can be achieved by deviation-proof policies), then determine the optimal equilibrium payoff based on the welfare function, and finally
construct the deviation-proof policy to achieve the optimal equilibrium payoff.

\subsection{Quantify The Set of Pareto Optimal Equilibrium Payoffs}
The first step in solving the design problem \eqref{eqn:PolicyDesignProblem} is to characterize the set of Pareto optimal equilibrium payoffs for the
dynamic spectrum sharing system. In particular, we are interested in the case when the SUs are impatient (their discount factor is strictly smaller
than $1$), as opposed to the asymptotic case when the SUs are arbitrarily patient (their discount factor goes to $1$) in
\cite{EtkinTse}\cite{WuWangLiu}\cite{FudenbergLevineMaskin94}. For repeated games with perfect monitoring, the characterization of Pareto optimal
equilibrium payoffs with impatient users is provided in \cite{XiaoMihaela_RepeatedGame}. Our result in
Theorem~\ref{theorem:CharacterizeEquilibriumPayoff} is the first one that analytically quantifies the set of Pareto optimal equilibrium payoffs for
repeated games with imperfect monitoring and impatient users.

For the spectrum sharing systems with strong multi-user interference, recall from Definition~\ref{definition:strong_interference} that the set of
feasible payoffs can be written as $\mathcal{V}=\mathrm{conv}\{(0,\ldots,0),
\mathbf{u}(\mathbf{\tilde{p}}^1),\ldots,\mathbf{u}(\mathbf{\tilde{p}}^N)\}$, and that its Pareto boundary is $\mathcal{B}=\{\mathbf{v}:\sum_{i=1}^N
v_i/\bar{v}_i=1,~v_i\geq0, \forall i\}$. Now we need to determine which portion of the Pareto boundary $\mathcal{B}$ can be achieved as equilibrium
payoffs (i.e. payoffs that can be achieved by deviation-proof policies).

Before stating Theorem~\ref{theorem:CharacterizeEquilibriumPayoff}, we define the \emph{benefit from deviation} as follows.
\begin{definition}[Benefit From Deviation]
We define SU $j$'s benefit from deviation from SU $i$'s payoff maximizing power profile $\mathbf{\tilde{p}}^i$ as
\begin{eqnarray}
b_{ij} = \max_{p_j\in \mathcal{P}_j, p_j\neq \tilde{p}_j^i}
\frac{\rho(y_0|\mathbf{\tilde{p}}^i)-\rho(y_0|p_j,\mathbf{\tilde{p}}_{-j}^i)}{u_j(p_j,\mathbf{\tilde{p}}_{-j}^i)/\bar{v}_j}.
\end{eqnarray}
\end{definition}
Our definition of the benefit from deviation results from two intuitions. First, whether there is a benefit from deviation should depend on whether
the deviation can be statistically detected. A deviation can be statistically detected only if
$\rho(y_0|\mathbf{\tilde{p}}^i)<\rho(y_0|p_j,\mathbf{\tilde{p}}_{-j}^i)$. This is because
$\rho(y_0|\mathbf{\tilde{p}}^i)<\rho(y_0|p_j,\mathbf{\tilde{p}}_{-j}^i)$ implies that the probability of sending the distress signal is larger when
the power profile is $(p_j,\mathbf{\tilde{p}}_{-j}^i)$, in which SU $j$ deviates from $\tilde{p}_j^i$ to $p_j$, than the corresponding probability
when the power profile is $\mathbf{\tilde{p}}^i$, in which SU $j$ does not deviate. Hence, it is statistically correct for the SUs to associate the
receipt of the distress signal $y_0$ with the event of deviation. Since $u_j(p_j,\mathbf{\tilde{p}}_{-j}^i)/\bar{v}_j$ is always larger than 0, the
benefit from deviation is negative if and only if $\rho(y_0|\mathbf{\tilde{p}}^i)<\rho(y_0|p_j,\mathbf{\tilde{p}}_{-j}^i)$. In other words, there is
no benefit but only cost from deviation if the deviation can be statistically identified by the distress signal.

Second, the benefit from deviation depends on how likely deviation can be detected (reflected by
$|\rho(y_0|\mathbf{\tilde{p}}^i)-\rho(y_0|p_j,\mathbf{\tilde{p}}_{-j}^i)|$), as well as how much a SU can gain from deviation (reflected by
$u_j(p_j,\mathbf{\tilde{p}}_{-j}^i)/\bar{v}_j$). Since $b_{ij}<0$, its absolute value $|b_{ij}|$ can be considered as the cost from deviation. The
cost from deviation $|b_{ij}|$ increases with $|\rho(y_0|\mathbf{\tilde{p}}^i)-\rho(y_0|p_j,\mathbf{\tilde{p}}_{-j}^i)|$, the likelihood that a
deviation is detected. In addition, $|b_{ij}|$ decreases with $u_j(p_j,\mathbf{\tilde{p}}_{-j}^i)/\bar{v}_j$, the payoff SU $j$ obtains from
deviation normalized by its maximum payoff.



Now we state Theorem~\ref{theorem:CharacterizeEquilibriumPayoff}, which analytically quantifies the set of Pareto optimal equilibrium payoffs.

\begin{theorem}\label{theorem:CharacterizeEquilibriumPayoff}
We can achieve the following set of Pareto optimal equilibrium payoffs
\begin{eqnarray}
\mathcal{B}_{\underline{\mu}} = \left\{\mathbf{v}:\sum_{i=1}^N \frac{v_i}{\bar{v}_i}=1,~\frac{v_i}{\bar{v}_i}\geq \underline{\mu}_i, \forall
i\in\mathcal{N}\right\},
\end{eqnarray}
where $\underline{\mu}_i \triangleq \max_{j\neq i} \frac{1-\rho(y_0|\mathbf{\tilde{p}}^j)}{-b_{ji}}$, if and only if first, the following two sets of
conditions are satisfied for all $i\in\mathcal{N}$ and for all $j\neq i$:
\begin{itemize}
\item Condition~1: benefit from deviation $b_{ij}<0$;
\item Condition~2: no incentive for SU $i$ to deviate:
\begin{eqnarray}
1-\frac{u_i(p_i,\mathbf{\tilde{p}}_{-i}^i)}{\bar{v}_i}+\sum_{j\neq i}
\frac{\rho(y_0|\mathbf{\tilde{p}}^i)-\rho(y_0|p_i,\mathbf{\tilde{p}}_{-i}^i)}{-b_{ij}}\geq0,\forall p_i, \nonumber
\end{eqnarray}
\end{itemize}
and second, the discount factor $\delta$ is larger than a threshold:
\begin{eqnarray}
\delta \geq \underline{\delta} \triangleq \frac{1}{1+\frac{1-\sum_{i\in\mathcal{N}} \underline{\mu}_i}{N-1+\sum_{i\in\mathcal{N}}\sum_{j\neq i}
(-\rho(y_0|\mathbf{\tilde{p}}^i)/b_{ij})}}.
\end{eqnarray}
\end{theorem}
\begin{IEEEproof}
We provide an outline of the proof here. Please refer to Appendix~\ref{proof:CharacterizeEquilibriumPayoff} for the complete proof.

The proof heavily replies on the concept of self-generating sets \cite{APS}. Simply put, a self-generating set, associated with a discount factor, is
a set in which every payoff is an PPE payoff under the associated discount factor \cite{APS}. Any self-generating set is associated with a minimum
discount factor; any discount factor larger than the minimum one can be associated with that self-generating set. The main contribution of the proof
is to find the largest self-generating set and the associated minimum discount factor. Since we focus on the Pareto optimal equilibrium payoffs, we
restrict to the self-generating sets on the Pareto boundary. This restriction allows us to obtain the analytical expression of the largest
self-generating set $\mathcal{B}_{\underline{\mu}}$. Meanwhile, the sufficient and necessary conditions for $\mathcal{B}_{\underline{\mu}}$ to be
self-generating are obtained.
\end{IEEEproof}

Theorem~\ref{theorem:CharacterizeEquilibriumPayoff} provides the sufficient and necessary conditions for the existence of Pareto optimal equilibrium
payoffs. Condition~1 (respectively, Condition~2) ensures that at the power profile $\mathbf{\tilde{p}}^i$, SU $j$ for any $j\neq i$ (respectively, SU
$i$) has no incentive to deviate. When the conditions are satisfied, Theorem~\ref{theorem:CharacterizeEquilibriumPayoff} quantifies the set of Pareto
optimal equilibrium payoffs $\mathcal{B}_{\underline{\mu}}$. We can choose any payoff in $\mathcal{B}_{\underline{\mu}}$ as the deviation-proof
operating point. Theorem~\ref{theorem:CharacterizeEquilibriumPayoff} also gives us the minimum discount factor under which any payoff in
$\mathcal{B}_{\underline{\mu}}$ is achievable. We can determine the maximum level of impatience the users can have in order to achieve any payoff in
$\mathcal{B}_{\underline{\mu}}$.

\begin{remark}
Note that we have not assumed the monotonicity of the payoff function $u_i$. If each SU's payoff function increases with its own transmit power, then
Condition~2 in Theorem~\ref{theorem:CharacterizeEquilibriumPayoff} holds true as long as Condition~1 is satisfied.
\end{remark}

\begin{remark}
Note also that the set of Pareto optimal equilibrium payoffs $\mathcal{B}_{\underline{\mu}}$ could be empty if $\underline{\mu}_i$ is large. More
precisely, $\mathcal{B}_{\underline{\mu}}$ is nonempty if and only if $\sum_{i\in\mathcal{N}} \underline{\mu}_i \leq 1$.
\end{remark}

\subsection{Determine The Optimal Operating Point}
Since we have identified the set of Pareto optimal equilibrium payoffs $\mathcal{B}_{\underline{\mu}}$, the problem of find the optimal operating
point that solves the policy design problem can be written as
\begin{eqnarray}\label{eqn:OptimalEquilibriumPayoff}
&\displaystyle\max_{\mathbf{v}}& W(v_1, \ldots, v_N) \\
&s.t.& (v_1/\bar{v}_1,\ldots,v_N/\bar{v}_N)\in \mathcal{B}_{\underline{\mu}}, \nonumber \\
&    & v_i\geq \gamma_i,~\forall i\in\mathcal{N}. \nonumber
\end{eqnarray}
The linear constraints in the above problem can be further simplified as $v_i\geq \max\{\underline{\mu}_i\cdot\bar{v}_i,\gamma_i\},~\forall
i\in\mathcal{N}$. Hence, we get the sufficient and necessary conditions under which the optimization problem \eqref{eqn:OptimalEquilibriumPayoff} is
feasible:
\begin{eqnarray}
\begin{array}{c}\sum_{i\in\mathcal{N}} \max\{\underline{\mu}_i,\gamma_i/\bar{v}_i\} \leq 1\end{array}.
\end{eqnarray}

The optimization problem \eqref{eqn:OptimalEquilibriumPayoff} is easy to solve when $W$ is a convex function in $(v_1,\ldots,v_N)$. For example, if
the objective function is the weighted sum of the users' payoffs, namely $W = \sum_{i=1}^N w_i v_i$, the solution can be obtained analytically as
$v_{i^*}^\star=(1-\sum_{j\neq i} \max\{\underline{\mu}_j,\gamma_j/\bar{v}_j\})\cdot\bar{v}_i$ for $i^*=\arg\max_{j\in\mathcal{N}} w_j \bar{v}_j$, and
$v_i^\star = \max\{\underline{\mu}_i,\gamma_i/\bar{v}_i\} \cdot \bar{v}_i$ for all $i\neq i^*$.

\subsection{Construct The Deviation-Proof Policy}
Given the optimal payoff $\mathbf{v}^\star\in\mathcal{B}_{\underline{\mu}}$, we can construct the deviation-proof policy that achieves the payoff
$\mathbf{v}^\star$. According to Definition~\ref{definition:strong_interference}, any payoff $\mathbf{v}^\star\in\mathcal{B}_{\underline{\mu}}$
should be achieved by alternating among $N$ operating points: $\mathbf{u}(\mathbf{\tilde{p}}^1),\ldots,\mathbf{u}(\mathbf{\tilde{p}}^N)$. Hence, the
deviation-proof policy $\sigma^*$ satisfies $\sigma^*(h^t)\in\{\mathbf{\tilde{p}}^1,\ldots,\mathbf{\tilde{p}}^N\}$ for any $t\geq0$ and for any
public history $h^t\in Y^t$. Since only one SU transmits in a time slot, the deviation-proof policy can also be regarded as a scheduling in a TDMA
fashion. By judiciously deciding which SU can transmit in each time slot, each SU $i$ receives a discounted expected average payoff $v_i^\star$ and
has no incentive to deviate from the policy. The deviation-proof policy can be implemented by each SU in a distributed manner. The algorithm run by
SU $i$ is described in the algorithm in Table~\ref{table:EquilibriumStrategy}.

The intuition of why the algorithm in Table~\ref{table:EquilibriumStrategy} works is as follows. At each time slot $t$, each SU $i$ calculates the
indices for all the SUs, $\alpha_i(t), \forall i\in\mathcal{N}$, where
\begin{eqnarray}
\alpha_j(t)=\frac{v_j(t)/\bar{v}_j-\underline{\mu}_j}{1-v_j(t)/\bar{v}_j+\sum_{k\neq j} (-\rho(y_0|\mathbf{\tilde{p}}^j)/b_{jk})},~\forall
j\in\mathcal{N}. \nonumber
\end{eqnarray}
The index $\alpha_i(t)$ measures SU $i$'s ``urgency'' to transmit at time slot $t$. The SU $i^*$ with the largest index $\alpha_{i^*}(t)=\max_i
\alpha_i(t)$ will transmit at time slot $t$. When no distress signal is received (which indicates no deviation), SU $i^*$'s index in the next time
slot is very likely to be small, in order to give the other SUs larger opportunities to transmit. However, when the distress signal is received
(which indicates deviation), they calculate the indices in a different way, such that SU $i^*$ still has a large index in the next time slot. Hence,
a SU may not have the incentive to deviate, because it will leads to a smaller opportunity to transmit in the future.

\begin{table}
\renewcommand{\arraystretch}{1.3}
\caption{The algorithm run by user $i$.} \label{table:EquilibriumStrategy} \centering
\begin{tabular}{l}
\hline
\textbf{Input:} The normalized target payoffs $\{v_i^\star/\bar{v}_i\}_{i\in\mathcal{N}}$ given by the LSS \\
\hline
\textbf{Initialization:} Set $t=0$, $v_j^\prime(0)=v_j^\star/\bar{v}_j$ for all $j\in\mathcal{N}$. \\
\textbf{repeat} \\
~~~~Calculates the index $\alpha_j(t) = \frac{v_j^\prime(t)-\underline{\mu}_j}{1-v_j^\prime(t)+\sum_{k\neq j} (-\rho(y_0|\mathbf{\tilde{p}}^j)/b_{jk})},\forall j$ \\
~~~~Finds the largest index $i^*\triangleq\arg\max_{j\in\mathcal{N}} \alpha_j(t)$ \\
~~~~\textbf{if} $i=i^*$ \textbf{then}\\
~~~~~~~~Transmits at the power level $\tilde{p}_i^i$ \\
~~~~\textbf{end~if} \\
~~~~Updates $v_j^\prime(t+1)$ for all $j\in\mathcal{N}$ as follows: \\
~~~~\textbf{if} No Distress Signal Received At Time Slot $t$ ($y^t=y_1$) \textbf{then} \\
~~~~~~~~$v_{i^*}^\prime(t+1)=\frac{1}{\delta}\cdot v_{i^*}^\prime(t)-(\frac{1}{\delta}-1)\cdot(1+\sum_{j\neq {i^*}} \frac{\rho(y_0|\mathbf{\tilde{p}}^{i^*})}{-b_{{i^*}j}})$ \\
~~~~~~~~$v_j^\prime(t+1)=\frac{1}{\delta}\cdot v_j^\prime(t)+(\frac{1}{\delta}-1)\cdot\frac{\rho(y_0|\mathbf{\tilde{p}}^{i^*})}{-b_{i^* j}},\forall j\in\mathcal{N}, j\neq i^*$ \\
~~~~\textbf{else} \\
~~~~~~~~$v_{i^*}^\prime(t+1)=\frac{1}{\delta}\cdot v_{i^*}^\prime(t)-(\frac{1}{\delta}-1)\cdot(1-\sum_{j\neq {i^*}} \frac{\rho(y_1|\mathbf{\tilde{p}}^{i^*})}{-b_{{i^*}j}})$ \\
~~~~~~~~$v_j^\prime(t+1)=\frac{1}{\delta}\cdot v_j^\prime(t)-(\frac{1}{\delta}-1)\cdot\frac{\rho(y_1|\mathbf{\tilde{p}}^{i^*})}{-b_{i^* j}},\forall j\in\mathcal{N}, j\neq i^*$ \\
~~~~\textbf{end~if} \\
~~~~$t\leftarrow t+1$ \\
\textbf{until} $\varnothing$ \\
\hline
\end{tabular}
\end{table}

Theorem~\ref{theorem:EquilibriumStrategy} ensures that if all the SUs run the algorithm in Table~\ref{table:EquilibriumStrategy} locally, they will
achieve the optimal operating point $\mathbf{v}^\star$, and will have no incentive to deviate.
\begin{theorem}\label{theorem:EquilibriumStrategy}
For any target payoff $\mathbf{v}^\star\in\mathcal{B}_{\underline{\mu}}$, and any discount factor $\delta\geq\underline{\delta}$, the strategy
generated by each user running the algorithm in Table~\ref{table:EquilibriumStrategy} is PPE and achieves $\mathbf{v}^\star$.
\end{theorem}
\begin{IEEEproof}
We provide an outline of the proof here. Please refer to Appendix~\ref{proof:EquilibriumStrategy} for the complete proof.

The key to the proof is to demonstrate that all the payoffs $\{v_i^\prime(t)\cdot \bar{v}_i\}_{i\in\mathcal{N}}, \forall t\geq0$ generated in the
algorithm in Table~\ref{table:EquilibriumStrategy} are in the self-generating set (the set of Pareto optimal equilibrium payoffs)
$\mathcal{B}_{\underline{\mu}}$.
\end{IEEEproof}

\subsection{Implementation Issues}

\begin{figure}
\centering
\includegraphics[width =3.5in]{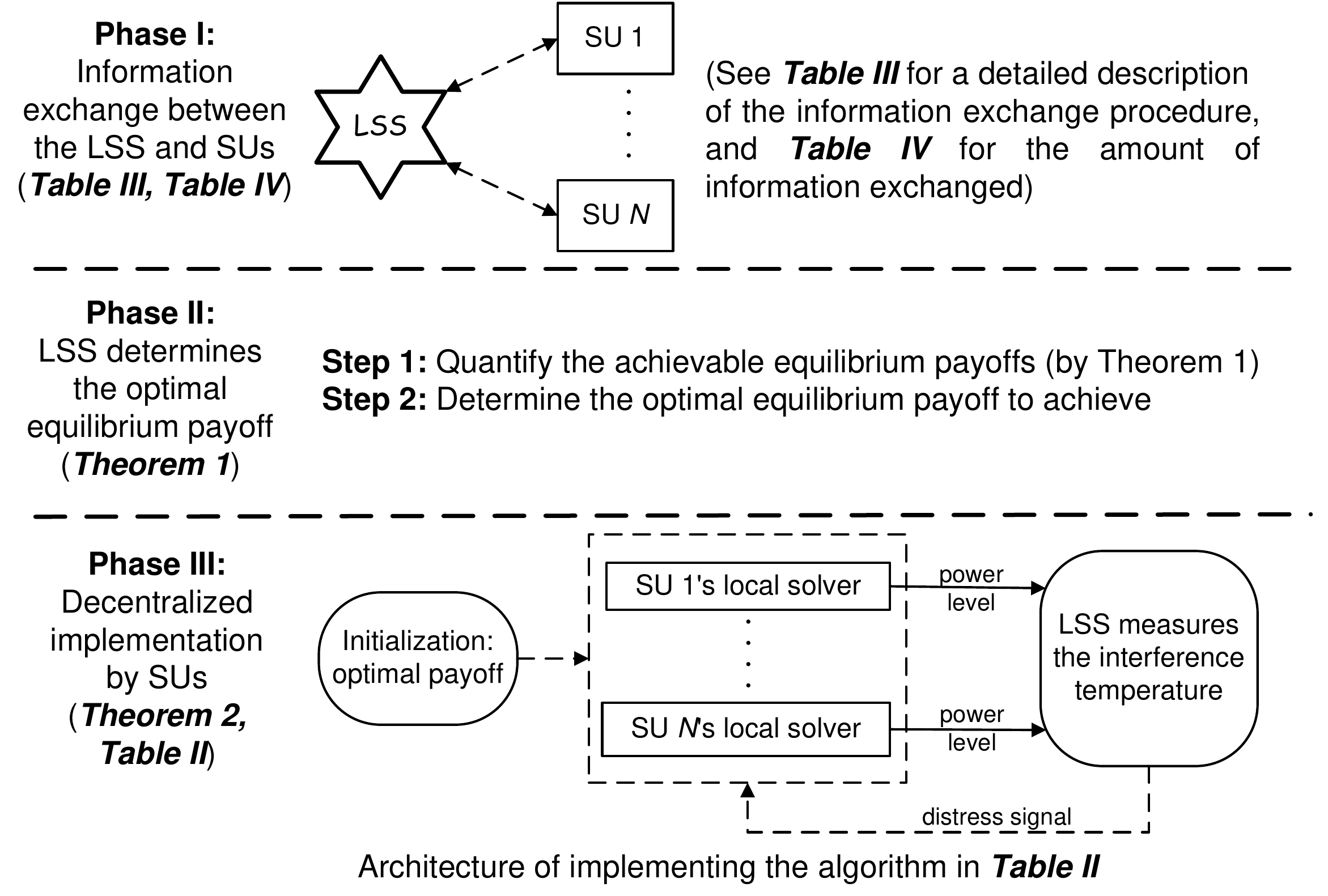}
\caption{Illustration of the implementation.} \label{fig:DesignFramework}
\end{figure}

\begin{table}
\renewcommand{\arraystretch}{1.1}
\caption{The Information Exchange Phase.} \label{table:InfoExchangePhase} \centering
\begin{tabular}{l|l}
\hline
Events & Information obtained \\
\hline \hline
SUs choose $\{\mathbf{\tilde{p}}^i\}_{i\in\mathcal{N}}$ & LSS: $\{\rho(y_0|\mathbf{\tilde{p}}^i)\}_{i\in\mathcal{N}}$ \\
\hline
SUs choose $(p_j^i,\mathbf{\tilde{p}}_{-j}^i),\forall j,p_j$ & LSS: $\rho(y_0|p_j^i,\mathbf{\tilde{p}}_{-j}^i),\forall j,p_j$\\
\hline \hline
LSS broadcasts & SU $i$: $\rho(y_0|\mathbf{\tilde{p}}^i),\rho(y_0|p_j^i,\mathbf{\tilde{p}}_{-j}^i)$ \\
\hline \hline
SUs broadcast & LSS, SUs: $b_{ij},\forall i,j\neq i$ \\
\hline
SUs send to LSS & LSS: $\{\bar{v}_i\}_{i\in\mathcal{N}}$ \\
\hline 
\end{tabular}
\end{table}

We discuss the implementation issues of our proposed design framework, which can be implemented in three phases as illustrated in
Fig.~\ref{fig:DesignFramework}. In Phase I, the LSS exchanges some information with the SUs following the procedure described in
Table~\ref{table:InfoExchangePhase}. In Phase II, using the information obtained in Phase I, the LSS quantifies the set of Pareto optimal equilibrium
payoffs, and solves the policy design problem for the optimal equilibrium payoff. Finally in Phase III, the LSS sends the optimal equilibrium payoff
to the SUs, as an input to each SU's decentralized algorithm of constructing the optimal deviation-proof policy.

\subsubsection{Overhead of information exchange} We briefly comment on the overhead of the information exchange in the proposed framework. First, the
information exchange in Phase I is necessary for the LSS to determine and for the SUs to achieve the optimal equilibrium payoff. A similar
information exchange phase is proposed in \cite{EtkinTse}\cite{WuWangLiu}\cite{CordeiroChallapali}--\cite{DeDomenico_MAC}. The information exchange
phase can be considered as a substitute for the convergence process needed by the algorithms in
\cite{HuangBerry_JSAC06}--\cite{GatsisMarquesGiannakis}\cite{SharmaTeneketzis_ToN}\cite{HuangBerry_06_Auction}\cite{SharmaTeneketzis_GameTheory}. In
the proposed policy, since the players implement the policy without any information exchange in Phase III, the only information exchange happen in
Phase I and at the end of Phase II (when the MU broadcasts the optimal equilibrium payoff). The information exchange method in our framework is
advantageous in that its duration and the amount of information to exchange are predetermined. On the other hand, the amount of information to
exchange in
\cite{HuangBerry_JSAC06}--\cite{GatsisMarquesGiannakis}\cite{SharmaTeneketzis_ToN}\cite{HuangBerry_06_Auction}\cite{SharmaTeneketzis_GameTheory} is
proportional to the convergence time of their algorithms, which are generally unbounded. We summarize the overhead of information exchange (measured
by the number of real numbers or pilot signals transmitted) in the related works in Table~\ref{table:InformationExchangeOverhead}.

\begin{table}
\renewcommand{\arraystretch}{1.1}
\caption{Comparison of the total amount of information exchanged.} \label{table:InformationExchangeOverhead} \centering
\begin{tabular}{|c|c|}
\hline
 & The total amount of information exchanged \\
\hline
\cite{HuangBerry_JSAC06}--\cite{GatsisMarquesGiannakis}\cite{HuangBerry_06_Auction} & $O(N)$ per iteration $\cdot$ $\#$ of iterations \\
\hline
\cite{SharmaTeneketzis_ToN}\cite{SharmaTeneketzis_GameTheory} & $O(N^2)$ per iteration $\cdot$ $\#$ of iterations \\
\hline
Proposed & $\sum_{i}\sum_{j\neq i} |A_j| + N^2 + 1$ \\
\hline
\end{tabular}
\end{table}

\subsubsection{Computational complexity} As we can see from Table~\ref{table:EquilibriumStrategy}, the computational complexity of each SU in constructing the optimal policy is very small.
At each period $t$, each SU only needs to compute $N$ indices $\{\alpha_j(t)\}_{j\in\mathcal{N}}$, and $N$ normalized payoffs
$\{v_j^\prime(t)\}_{j\in\mathcal{N}}$, all of which can be calculated by analytical expressions. In addition, although the original definition of the
strategy requires each SU to memorize the entire history of measurement outcomes, in the actual implementation, each SU only needs to know the
current measurement outcome and memorize $N$ normalized payoffs $\{v_j^\prime(t)\}_{j\in\mathcal{N}}$.

\section{Simulation Results}\label{sec:Simulation}
In this section, we demonstrate the performance gain of our spectrum sharing policy over existing policies, and validate our theoretical analysis
through numerical results. Throughout this section, we use the following system parameters by default unless we change some of them explicitly. The
noise powers at all the SUs' receivers are normalized as $0$ dB. The maximum transmit powers of all the SUs are $10$ dB, $\forall i$. For simplicity,
we assume that the direct channel gains have the same distribution $g_{ii}\thicksim\mathcal{CN}(0,1), \forall i$, and the cross channel gains have
the same distribution $g_{ij}\thicksim\mathcal{CN}(0,\beta), \forall i\neq j$, where $\beta$ is defined as the \emph{cross interference level}. The
channel gain from each SU to the LSS also satisfies $g_{i0}\thicksim\mathcal{CN}(0,1), \forall i$. The IT limit set by the PU is $\bar{I}=10$ dB. The
measurement error $\varepsilon$ is Gaussian distributed with zeros mean and variance $0.1$. The maximum false alarm probability is
$\bar{\Gamma}=10\%$. The SUs' payoffs are their throughput as in \eqref{eqn:Throughput}. The welfare function is the average payoff, i.e.
$W=\sum_{i=1}^N \frac{1}{N} U_i$. The minimum payoff guarantee is $10\%$ of the maximum achievable payoff, i.e. $\gamma_i=0.1\cdot \bar{v}_i, \forall
i$.
\subsection{Performance Evaluation}
\subsubsection{Comparison with policies with constant power levels}
We first compare the performance of the proposed policy with that of the optimal policy with constant power levels. The optimal policy with constant
power levels (or ``the optimal stationary policy'') is the solution to the modified version of the design problem \eqref{eqn:PolicyDesignProblem}.
First, we add an additional constraint that the power profile is constant, namely $\bm{\sigma}(h^t)=\mathbf{p}^\star$ for all $t\geq0$ and for all
$h^t\in Y^t$. Second, we drop the incentive constraint that $\bm{\sigma}$ is PPE from \eqref{eqn:PolicyDesignProblem}. Hence, the performance of the
optimal stationary policy is the best that can be achieved by existing stationary policies \cite{HuangBerry_JSAC06}--\cite{SorooshyariTanChiang}, and
is an upper bound for the deviation-proof stationary policies \cite{HuangBerry_06_Auction}--\cite{XiaovanderSchaar_PowerControl}.

\begin{figure}
\centering
\includegraphics[width =3.5in]{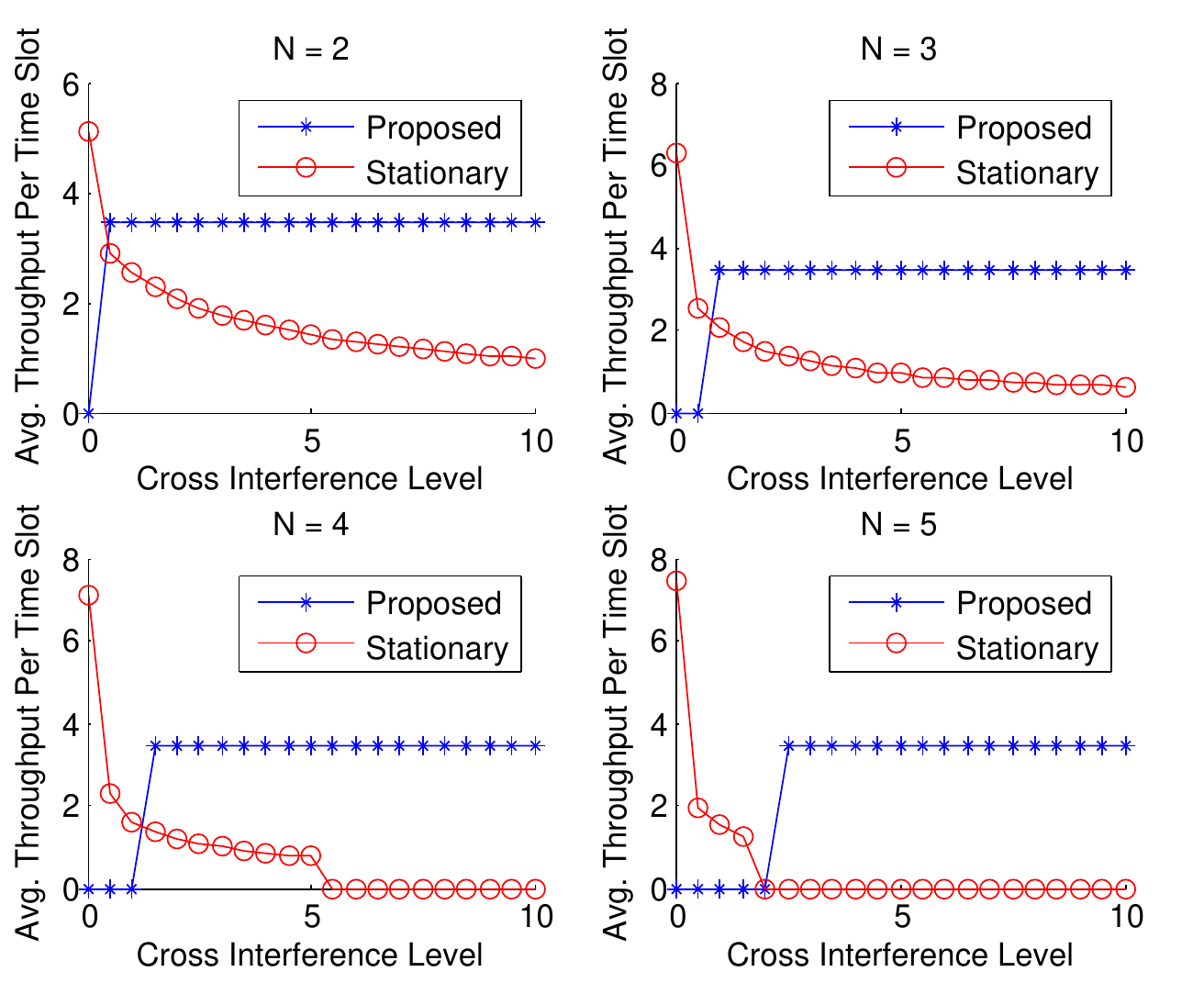}
\caption{Performance comparison of the proposed policy and the optimal policy with constant power levels (`stationary' in the legend) under different
numbers of users and different cross interference levels. A zero average throughput indicates that there exists no feasible policy that satisfies all
the constraints in the policy design problem.} \label{fig:Comparison_Stationary_CrossInterferenceLevel}
\end{figure}

In Fig.~\ref{fig:Comparison_Stationary_CrossInterferenceLevel}, we compare the performance of the proposed policy and that of the optimal stationary
policy under different cross interference levels and different numbers of SUs. As expected, the proposed policy outperforms the optimal stationary
policy in medium to high cross interference levels (approximately when $\beta\geq1$). In the cases of high cross interference levels ($\beta\geq2$)
and many users ($N=5$), the stationary policy fails to meet the minimum payoff guarantees due to strong interference (indicated by zero average
throughput in the figure). On the other hand, the desirable feature of the proposed policy is that the average throughput does not decrease with the
increase of the cross interference level, because SUs transmit in a TDMA fashion. For the same reason, the average throughput does not change with
the number of SUs.

Note that the proposed policy is infeasible (zero average throughput) when the cross interference level is very small. This is because it cannot be
deviation-proof in this scenario. When the interference level is very small, SU $j$ can deviate from $\mathbf{\tilde{p}}^i$ and receives a high
reward $u_j(p_j,\mathbf{\tilde{p}}_{-j}^i)$ because the interference from SU $i$, $\tilde{p}_i^i g_{ij}$, is small. Hence, the benefit of deviation
$b_{ij}$ is large, and the deviation is inevitable. This observation leads to an efficient way for the LSS to check the cross interference level
without knowing the channel gains. If the proposed policy is infeasible, the LSS knows that the cross interference level is low, and can switch to
stationary policies.

\subsubsection{Comparison with ``punish-forgive'' policies proposed under perfect monitoring}
We also compare the proposed policy with existing policies designed under the assumption of perfect monitoring
\cite{EtkinTse}--\cite{XiaoMihaela_RepeatedGame}. Specifically, we consider the ``punish-forgive'' policy in
\cite{EtkinTse}--\cite{XiaoMihaela_RepeatedGame}, which requires SUs to switch to the punishment phase of $L$ time slots once a deviation is
detected. In the punishment phase, all the SUs transmit at the maximum power levels to create high interference to the deviator\footnote{Note that
all the SUs transmitting at the maximum power levels. For the punish-forgive policy \cite{EtkinTse}--\cite{XiaoMihaela_RepeatedGame}, we allow the
violation of the IT constraint in the punishment phase. Note that the IT constraint is never violated in the proposed policy.}. A special case of the
punish-forgive policy when the punishment length $L=\infty$ \cite{EtkinTse} is the celebrated ``grim-trigger'' strategy in game theory literature
\cite{MailathSamuelson}. As discussed before, the punish-forgive policy works well if the SUs can perfectly monitor the individual power levels of
all the SUs, because in this case, the punishment serves as a threat and will never be carried out in the equilibrium. However, when the SUs have
imperfect monitoring ability, the punishment will be carried out with some positive probability, which decreases all the SUs' average payoffs.

\begin{figure}
\centering
\includegraphics[width =3.5in]{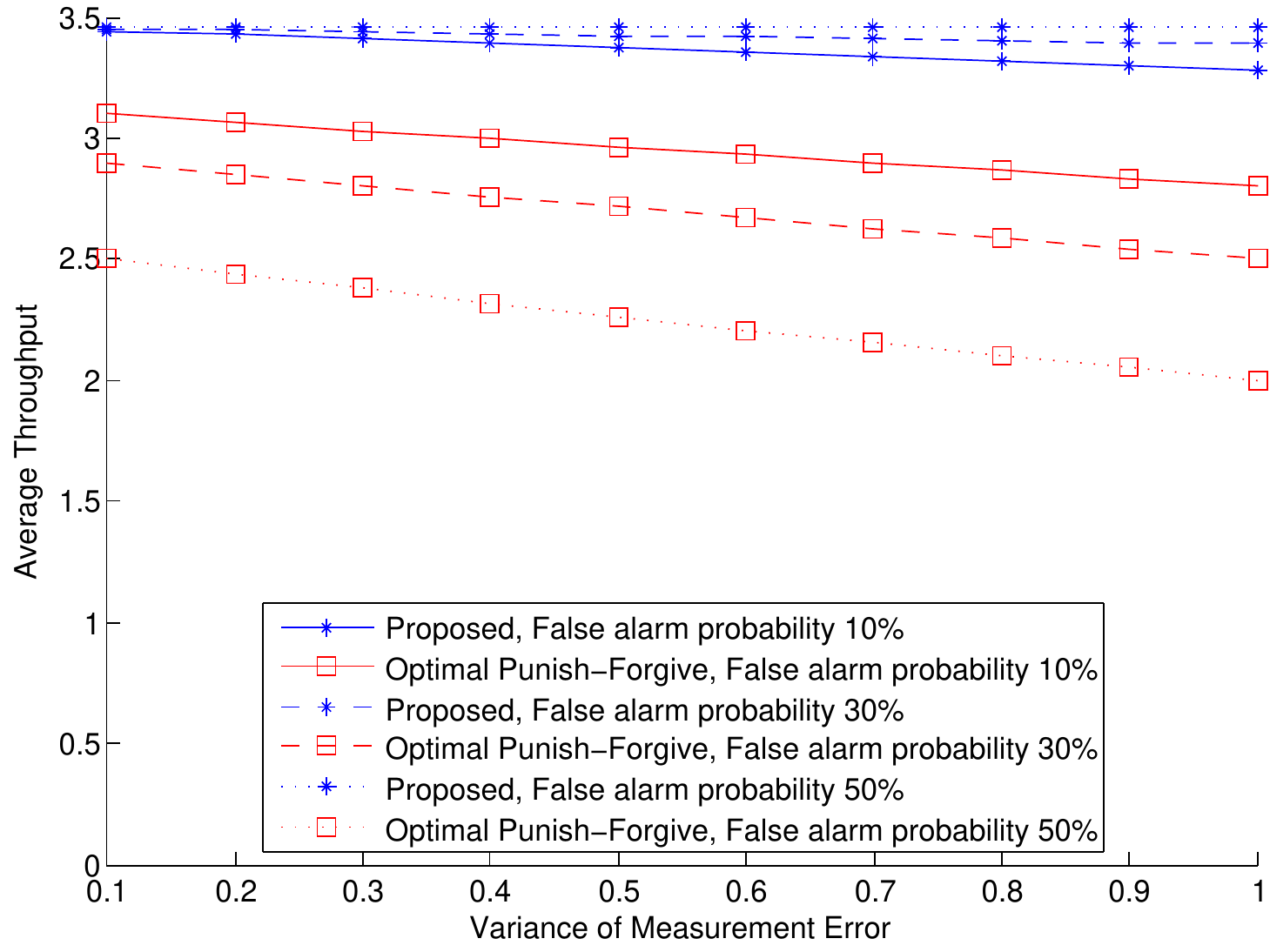}
\caption{Performance comparison of the proposed policy and the punish-forgive policy with the optimal punishment length under different error
variances and different false alarm probabilities.} \label{fig:Comparison_PunishForgive_Variance_FalseAlarmProbability}
\end{figure}

Fig.~\ref{fig:Comparison_PunishForgive_Variance_FalseAlarmProbability} shows that the proposed policy outperforms the punish-forgive policies under
different variances of measurement errors and different false alarm probabilities. For each combination of the error variance and the false alarm
probability, we choose the punish-forgive policy with the optimal punishment length. The performance of punish-forgive polices degrades with the
increase of the error variance and the false alarm probability, because of the increasing probability of mistakenly triggered punishments. Some
interesting observation on how the performance of the proposed policy changes with the error variance and the false alarm probability is explained in
details in the following subsections.

\subsection{Impacts of Variances of Measurement Errors}
\begin{figure}
\centering
\includegraphics[width =3.5in]{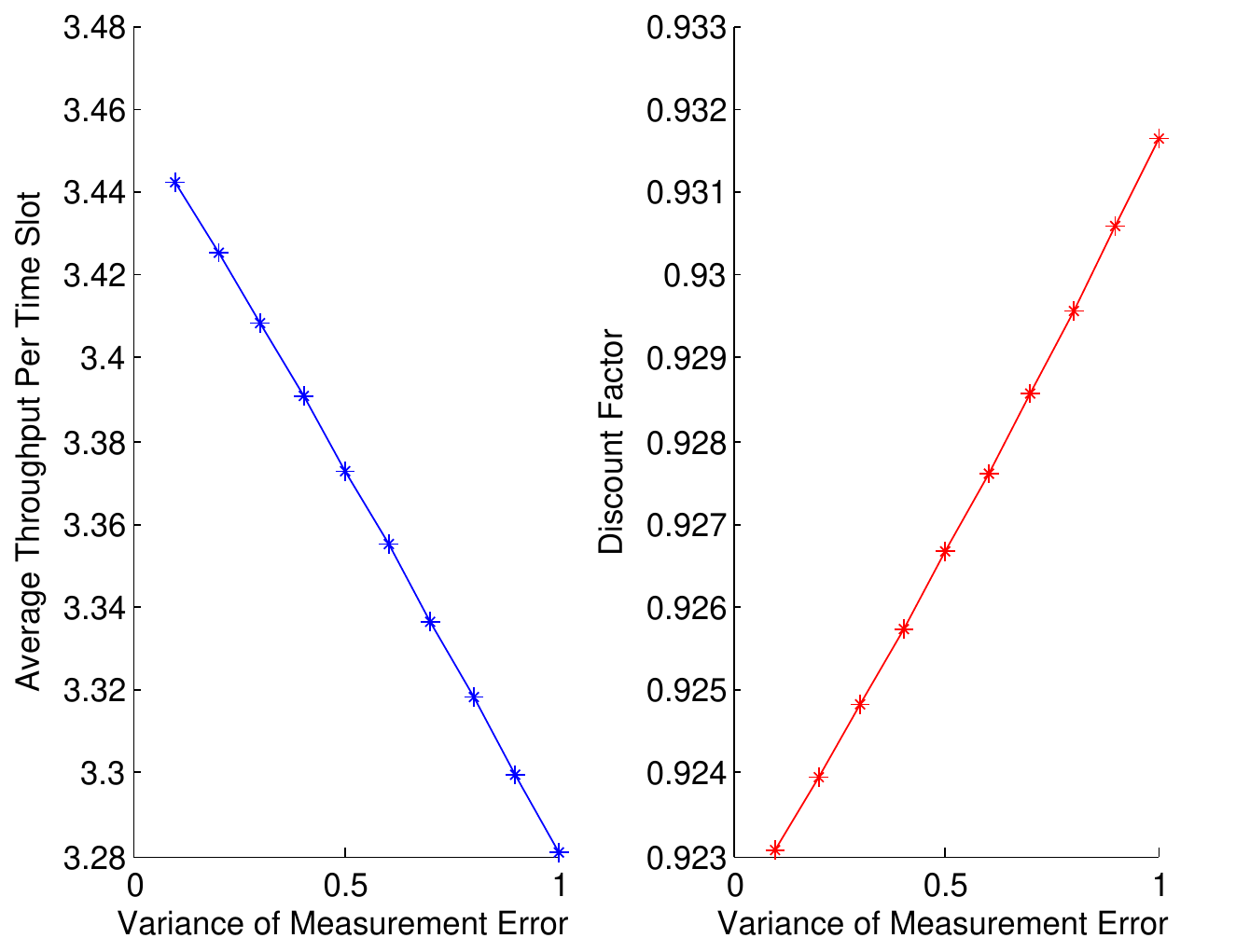}
\caption{The impact of the variance of the measurement error on the performance of the proposed policy and the minimum discount factor required under
which the proposed policy is deviation-proof.} \label{fig:Property_Variance}
\end{figure}

Fig.~\ref{fig:Property_Variance} shows that with the increase of the variance of measurement errors, the average throughput decreases, and the SUs'
patience (the discount factor) required to achieve Pareto optimal equilibrium payoffs increases. First, when the error variance increases, the
intermediate IT limit $I$ must decrease to maintain the constraint on the false alarm probability. The decrease of $I$ leads to the decrease of SUs'
maximum transmit power levels allowed, which results in the decrease of the average throughput. Another impact of the increase in the error variance
is that $\rho(y_0|p_j,\mathbf{\tilde{p}}_{-j}^i)=\int_{x>\bar{I}-p_j h_{j0}-\tilde{p}_i^i h_{i0}} f_{\varepsilon}(x) dx$ increases, which leads to
the increase of benefit of deviation $b_{ij}$. Hence, the minimum discount factor $\underline{\delta}$ increases according to
Theorem~\ref{theorem:CharacterizeEquilibriumPayoff}.

\begin{figure}
\centering
\includegraphics[width =3.5in]{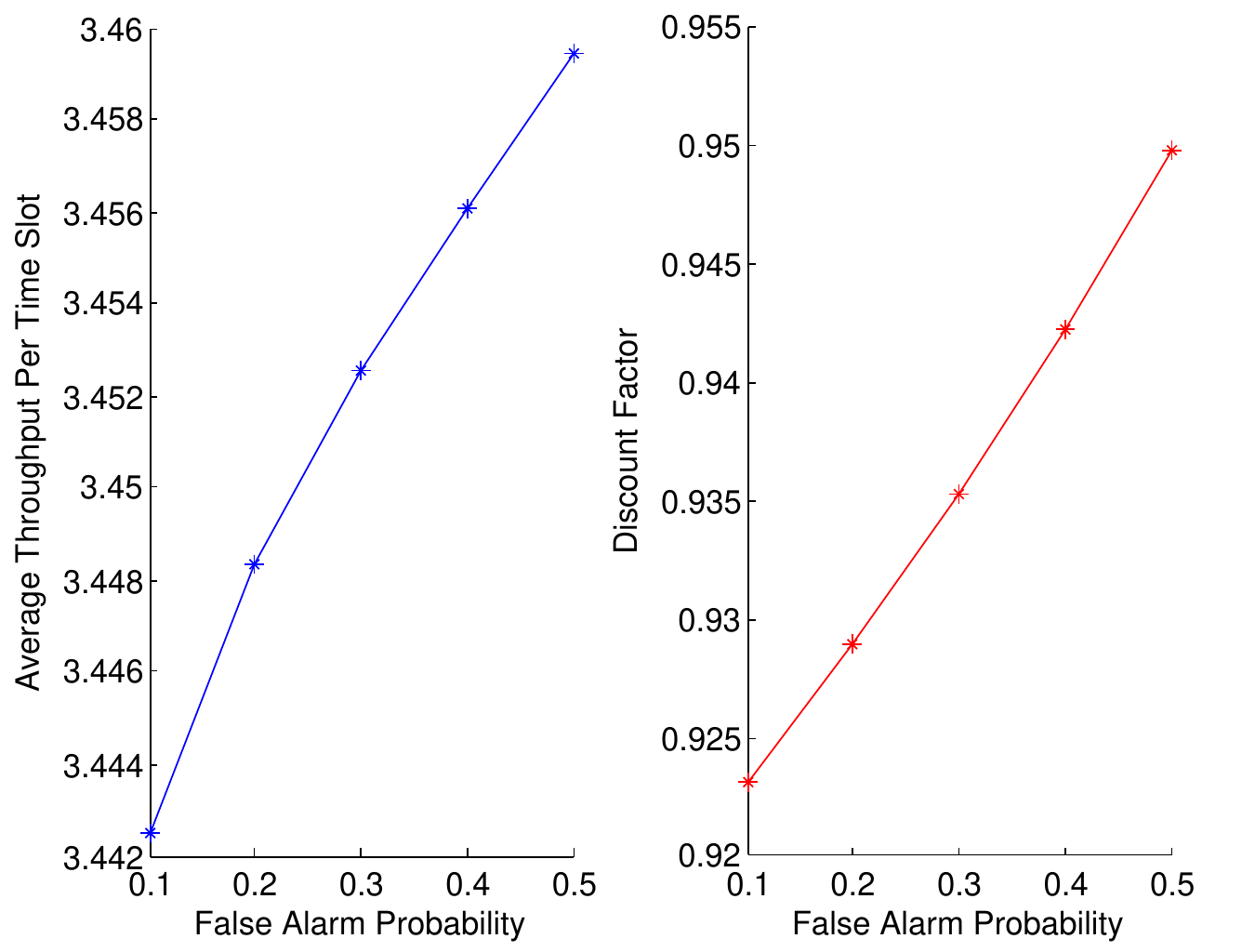}
\caption{The impact of the false alarm probability on the performance of the proposed policy and the minimum discount factor required under which the
proposed policy is deviation-proof.} \label{fig:Property_FalseAlarmProbability}
\end{figure}

\subsection{Impacts of Constraints on The False Alarm Probability}
Fig.~\ref{fig:Property_FalseAlarmProbability} shows that with the increase of the false alarm probability limit $\bar{\Gamma}$, both the average
throughput and the users' patience (the discount factor) required to achieve Pareto optimal equilibrium payoffs increase. First, with an increased
false alarm probability limit, the intermediate IT limit $I$ can increase, which leads to an increase of the SUs' maximum transmit power levels and
thus an increase of the users' throughput. Meanwhile, since
\begin{eqnarray}
\rho(y_0|\mathbf{\tilde{p}}^i)-\rho(y_0|p_j,\mathbf{\tilde{p}}_{-j}^i) = -\int_{\bar{I}-I-h_{0j}p_j}^{\bar{I}-I} f_{\varepsilon}(x) dx \nonumber
\end{eqnarray}
increases when $I$ increases, the benefit of deviation $b_{ij}$ increases. This leads to an increase of the minimum discount factor.

This observation indicates an interesting design tradeoff. On one hand, a smaller false alarm probability can reduce the overhead of sending distress
signals, and can also relax the requirement on SUs' patience. On the other hand, a larger false alarm probability can increase the average
throughput, such that the spectrum efficiency or the revenue can increase. Our theoretical results characterize such a tradeoff, which can be used to
choose the optimal intermediate IT limit $I$.

\section{Conclusion}\label{sec:Conclusion}
In this paper, we studied power control in dynamic spectrum sharing among SUs under the interference temperature constraint, and proposed a dynamic
spectrum sharing policy that allows SUs to transmit in a TDMA fashion. The proposed policy can achieve Pareto optimal operating points that are not
achievable under existing spectrum sharing policies with constant power levels. The proposed policy is amenable to distributed implementation and is
deviation-proof, in that the SUs are in their self-interests (i.e. maximizing their own QoS) to follow the policy. The proposed policy can achieve
Pareto optimality even when the SUs have limited and imperfect monitoring ability: they only observe distress signals that erroneously indicate the
violation of the interference temperature constraint. Simulation results validate our analytical results on the policy design and demonstrate the
performance gains enabled by the proposed policy.

\appendices
\section{Proof of Theorem~\ref{theorem:CharacterizeEquilibriumPayoff}}\label{proof:CharacterizeEquilibriumPayoff}
The proof culminates in the demonstration that under certain conditions, a set of Pareto optimal payoffs can be a \emph{self-generating} set. Then
according to \cite[Proposition~7.3.1]{MailathSamuelson}\cite{APS}, all the payoffs in the set are equilibrium payoffs. More specifically, we derive
the sufficient and necessary conditions (i.e. Conditions~1-3 in Theorem~\ref{theorem:CharacterizeEquilibriumPayoff}) under which a subset of Pareto
optimal payoffs is a self-generating set, and find the largest subset of Pareto optimal payoffs that can be self-generating (i.e.
$\mathcal{B}_{\bm{\underline{\mu}}}$ defined in Theorem~\ref{theorem:CharacterizeEquilibriumPayoff}).

\subsection{Preliminaries on Self-generating Sets}
We first provide some background knowledge related to the self-generating sets. Similar to Markov decision processes (MDP's), when we analyze the
game, we can decompose the average payoff into the current payoff and the continuation payoff (i.e. the average payoff starting from the next time
slot). However, there are two key differences between the decomposition in a game and that in a MDP. First, there are \emph{multiple} users in a
game, as opposed to MDP's in which there is usually only one user. Second, the incentive compatibility constraints, which are not present in a MDP,
need to be considered in a game. Hence, the \emph{decomposability} in a game is defined as follows
\cite[Definition~7.3.2]{MailathSamuelson}\cite{APS}.\footnote{For the ease of reference, we duplicate the definition in
\cite[Definition~7.3.2]{MailathSamuelson} here.}
\begin{definition}[Decomposability]\label{definition:decomposability}
A payoff $\mathbf{v}\in\mathbb{R}^N$ is \emph{decomposable} on a set $\mathcal{W}\subseteq\mathbb{R}^N$ with respect to discount factor $\delta$ and
(pure) action profile $\mathbf{p}$, if there exists a mapping $\bm{\gamma}:Y\rightarrow\mathcal{W}$, such that for all $i\in\mathcal{N}$, we have
\begin{eqnarray}
v_i &=& (1-\delta)\cdot u_i(\mathbf{p}) + \delta \cdot \sum_{y\in Y} \gamma_i(y)\rho(y|\mathbf{p}) \\
    &\geq& (1-\delta)\cdot u_i(p_i^\prime,\mathbf{p}_{-i}) + \delta \cdot \sum_{y\in Y} \gamma_i(y)\rho(y|p_i^\prime,\mathbf{p}_{-i}),~\forall p_i^\prime\in\mathcal{P}_i.
\end{eqnarray}
A payoff $\mathbf{v}$ is decomposable on a set $\mathcal{W}$ with respect to discount factor $\delta$, if there exists an action profile
$\mathbf{p}$, such that $\mathbf{v}$ is decomposable on a set $\mathcal{W}$ with respect to discount factor $\delta$ and action profile $\mathbf{p}$.
\end{definition}

In the above definition, we can see that each user $i$'s payoff $v_i$ is decomposed into the current payoff $u_i(\mathbf{p})$ and the expected
continuation payoff $\sum_{y\in Y} \gamma_i(y)\rho(y|\mathbf{p})$, which specifies the continuation payoff $\gamma_i(y)$ starting from the next
period given the signal $y$. Importantly, the decomposition needs to be \emph{incentive compatible}, in the sense that each user $i$ cannot choose a
different action $p_i^\prime$ to improve the average payoff. For convenience, we write $\mathscr{D}(\mathcal{W};\delta,\mathbf{p})$ as the set of
payoffs that can be decomposed on set $\mathcal{W}$ with respect to discount factor $\delta$ and action profile $\mathbf{p}$, namely
\begin{eqnarray}
\mathscr{D}(\mathcal{W};\delta,\mathbf{p}) =
\{\mathbf{v}\in\mathbb{R}^N:~\mathbf{v}~\mathrm{is~decomposable~on~set}~\mathcal{W}~\mathrm{with~respect~to}~\delta~\mathrm{and}~\mathbf{p}.\}
\end{eqnarray}
Similarly, we write $\mathscr{D}(\mathcal{W};\delta)\triangleq\cup_{\mathbf{p}\in\mathcal{P}} \mathscr{D}(\mathcal{W};\delta,\mathbf{p})$ as the set
of payoffs that can be decomposed on set $\mathcal{W}$ with respect to discount factor $\delta$.

A self-generating set is a set $\mathcal{W}$, in which every payoff $\mathbf{v}\in\mathcal{W}$ is decomposable on the set $\mathcal{W}$ itself. The
formal definition is as follows \cite[Definition~7.3.4]{MailathSamuelson}\cite{APS}.
\begin{definition}[Self-generating Sets]\label{definition:self-generating}
A set $\mathcal{W}$ is self-generating under discount factor $\delta$, if $\mathcal{W}\subseteq\mathscr{D}(\mathcal{W};\delta)$.
\end{definition}

The self-generating sets play an important role in repeated game theory, because every payoff in a self-generating set is an equilibrium payoff. We
restate this important result formally in the following lemma \cite[Proposition~7.3.1]{MailathSamuelson}\cite{APS}.
\begin{lemma}[Self-generation]
For any bounded set $\mathcal{W}\subset\mathbb{R}^N$, if $\mathcal{W}$ is self-generating, then every payoff in $\mathcal{W}$ is an equilibrium
payoff of the repeated game.
\end{lemma}

\subsection{Outline of The Proof}
In the above subsection, we have summarized some important results related to self-generation in repeated game theory. Now we outline the proof of
Theorem~\ref{theorem:CharacterizeEquilibriumPayoff}.

Recall that due to Definition~\ref{definition:strong_interference}, the Pareto boundary of the considered repeated game is
$$
\mathcal{B}=\left\{\mathbf{v}:\sum_{i\in\mathcal{N}} \frac{v_i}{\bar{v}}=1,~v_i\geq0,~\forall i\in\mathcal{N}\right\}.
$$
Consider a subset of the Pareto boundary
\begin{eqnarray}
\mathcal{B}_{\bm{\mu}}\triangleq \left\{\mathbf{v}:\sum_{i\in\mathcal{N}} \frac{v_i}{\bar{v}}=1,~\frac{v_i}{\bar{v}}\geq\mu_i,~\forall
i\in\mathcal{N}\right\},
\end{eqnarray}
where $\mu_i\geq0$ for all $i\in\mathcal{N}$. Our focus is to show that under certain conditions, the subset of the Pareto boundary
$\mathcal{B}_{\bm{\mu}}$ can be a self-generating set, which means that every Pareto optimal payoff in $\mathcal{B}_{\bm{\mu}}$ can be an equilibrium
payoff. In the next subsection, we derive the necessary conditions if $\mathcal{B}_{\bm{\mu}}$ is self-generating. These necessary conditions lead to
Conditions~1-3 in Theorem~\ref{theorem:CharacterizeEquilibriumPayoff}. A byproduct of the first necessary condition are the constraints on the
boundary $\bm{\mu}$ of the self-generating sets $\mathcal{B}_{\bm{\mu}}$ (i.e. the lower bound $\bm{\underline{\mu}}$ of $\bm{\mu}$ in
Theorem~\ref{theorem:CharacterizeEquilibriumPayoff}), which leads to the characterization of the largest possible self-generating set
$\mathcal{B}_{\bm{\underline{\mu}}}$. In the final subsection, we show that these necessary conditions are also sufficient for
$\mathcal{B}_{\bm{\mu}}$ to be self-generating.


\subsection{Necessary Conditions For a Set of Pareto Optimal Payoffs To Be Self-generating}
Suppose that $\mathcal{B}_{\bm{\mu}}$ is self-generating. Then for any payoff $\mathbf{v}\in\mathcal{B}_{\bm{\mu}}$, there exists an action profile
$\mathbf{p}$ and a mapping $\bm{\gamma}:Y\rightarrow\mathcal{B}_{\bm{\mu}}$, such that for all $i\in\mathcal{N}$, we have
\begin{eqnarray}
v_i &=& (1-\delta)\cdot u_i(\mathbf{p}) + \delta \cdot \sum_{y\in Y} \gamma_i(y)\rho(y|\mathbf{p}) \\
    &\geq& (1-\delta)\cdot u_i(p_i^\prime,\mathbf{p}_{-i}) + \delta \cdot \sum_{y\in Y} \gamma_i(y)\rho(y|p_i^\prime,\mathbf{p}_{-i}),~\forall p_i^\prime\in\mathcal{P}_i.
\end{eqnarray}
The first observation is that the action profile $\mathbf{p}$ that decomposes a Pareto optimal payoff $\mathbf{v}\in\mathcal{B}_{\bm{\mu}}$ must be a
payoff-maximizing action profile for a certain user. In other words, $\mathbf{p}\in\{\mathbf{\tilde{p}}^1,\ldots,\mathbf{\tilde{p}}^N\}$. This is
because the average payoff $\mathbf{v}$ and the continuation payoffs $\bm{\gamma}(y),\forall y\in Y,$ are all on the Pareto boundary $\mathcal{B}$.
In other words, $\sum_{i\in\mathcal{N}} v_i/\bar{v}_i=1$ and $\sum_{i\in\mathcal{N}} \gamma_i(y)/\bar{v}_i=1,\forall y\in Y$. Since the average
payoff is the convex combination of the current payoff and the expected continuation payoff, the current payoff must also lie on the Pareto boundary,
i.e. $\sum_{i\in\mathcal{N}} u_i(\mathbf{p})/\bar{v}_i=1$. According to Definition~\ref{definition:strong_interference}, the only action profiles
that lie on the Pareto boundary are $\mathbf{\tilde{p}}^1,\ldots,\mathbf{\tilde{p}}^N$.

Based on the above observation, we have $\mathscr{D}(\mathcal{W};\delta)=\cup_{i\in\mathcal{N}}
\mathscr{D}(\mathcal{W};\delta,\mathbf{\tilde{p}}^i)$. Suppose that a payoff $\mathbf{v}\in\mathcal{B}_{\bm{\mu}}$ is decomposed by
$\mathbf{\tilde{p}}^i$, namely $\mathbf{v}\in\mathscr{D}(\mathcal{W};\delta,\mathbf{\tilde{p}}^i)$. Using the facts that
$u_i(\mathbf{\tilde{p}}^i)=\bar{v}_i$ and $u_j(\mathbf{\tilde{p}}^i)=0,\forall j\neq i$, we have
\begin{eqnarray}\label{eqn:IC_active}
v_i &=& (1-\delta)\cdot \bar{v}_i + \delta \cdot \sum_{y\in Y} \gamma_i(y)\rho(y|\mathbf{\tilde{p}}^i) \\
    &\geq& (1-\delta)\cdot u_i(p_i,\mathbf{\tilde{p}}^i_{-i}) + \delta \cdot \sum_{y\in Y} \gamma_i(y)\rho(y|p_i,\mathbf{\tilde{p}}^i_{-i}),~\forall
    p_i\in\mathcal{P}_i, \nonumber
\end{eqnarray}
and for all $j\neq i$,
\begin{eqnarray}\label{eqn:IC_inactive}
v_j &=& \delta \cdot \sum_{y\in Y} \gamma_j(y)\rho(y|\mathbf{\tilde{p}}^i) \\
    &\geq& (1-\delta)\cdot u_j(p_j,\mathbf{\tilde{p}}^i_{-j}) + \delta \cdot \sum_{y\in Y} \gamma_j(y)\rho(y|p_j,\mathbf{\tilde{p}}^i_{-j}),~\forall
    p_j\in\mathcal{P}_j. \nonumber
\end{eqnarray}
Since user $j\neq i$ chooses $\tilde{p}_j^i=0$ in action profile $\mathbf{\tilde{p}}^i$, we say that under action profile $\mathbf{\tilde{p}}^i$,
user $i$ is the active user and user $j\neq i$ is an inactive user.

Next, we show that the incentive compatibility constraints for inactive users and the active user imply Condition 1 and Condition 2 of
Theorem~\ref{theorem:CharacterizeEquilibriumPayoff}, respectively. The incentive constraints for inactive users also give us constraints on the
boundary $\bm{\mu}$ of $\mathcal{B}_{\bm{\mu}}$. In addition, to make sure that $\bm{\gamma}(y)\in\mathcal{B}_{\bm{\mu}},\forall y$, the discount
factor should satisfy Condition 3 of Theorem~\ref{theorem:CharacterizeEquilibriumPayoff}.

\subsubsection{Incentive Constraints For Inactive Users}
We examine the incentive compatibility constraint for an inactive users $j\neq i$ in \eqref{eqn:IC_inactive}, which will lead to the first necessary
condition. First, since $u_j(p_j,\mathbf{\tilde{p}}^i_{-j})>0,\forall p_j>0$, for the inequality in \eqref{eqn:IC_inactive} to hold, we must have
$\sum_{y\in Y} \gamma_j(y)\rho(y|\mathbf{\tilde{p}}^i)>\sum_{y\in Y} \gamma_j(y)\rho(y|p_j,\mathbf{\tilde{p}}^i_{-j})$, which is equivalent to
\begin{eqnarray}
\left[\rho(y_0|\mathbf{\tilde{p}}^i)-\rho(y_0|p_j,\mathbf{\tilde{p}}^i_{-j})\right]\cdot (\gamma_j(y_0)-\gamma_j(y_1)) > 0,~\forall p_j>0.
\end{eqnarray}
Note that the probability of receiving distress signals given action profile $(p_j,\mathbf{\tilde{p}}^i_{-j})$ is no smaller than the probability
given $\mathbf{\tilde{p}}^i$, because
\begin{eqnarray}
\rho(y_0|p_j,\mathbf{\tilde{p}}^i_{-j})-\rho(y_0|\mathbf{\tilde{p}}^i) = \int_{\bar{I}-\tilde{p}_i^i g_{i0}-p_j g_{j0}}^{\bar{I}-\tilde{p}_i^i
g_{i0}} f_{\varepsilon}(x) dx \geq 0.
\end{eqnarray}
Since $\rho(y_0|p_j,\mathbf{\tilde{p}}^i_{-j})\geq\rho(y_0|\mathbf{\tilde{p}}^i)$, we must have $\gamma_j(y_1)>\gamma_j(y_0)$. This requirement is
intuitive: we should set a lower continuation payoff following the distress signal $y_0$ in order to deter user $j\neq i$ from deviating from
$\mathbf{\tilde{p}}^i$.

From the equality constraint in \eqref{eqn:IC_inactive}, we have
\begin{eqnarray}
\delta = \frac{v_j}{\sum_{y\in Y} \gamma_j(y)\rho(y|\mathbf{\tilde{p}}^i)}.
\end{eqnarray}
Plugging in the above expression of $\delta$, we can eliminate discount factor $\delta$ in the inequality of \eqref{eqn:IC_inactive} and obtain an
equivalent inequality as follows
\begin{eqnarray}
\sum_{y\in Y}
\gamma_j(y)\left[\left(1-\frac{v_j}{u_j(p_j,\mathbf{\tilde{p}}^i_{-j})}\right)\rho(y|\mathbf{\tilde{p}}^i)+\frac{v_j}{u_j(p_j,\mathbf{\tilde{p}}^i_{-j})}\rho(y|p_j,\mathbf{\tilde{p}}^i_{-j}))\right]
\leq v_j,~\forall p_j\neq \tilde{p}_j^i.
\end{eqnarray}
For notational simplicity, we write the coefficient of $\gamma_j(y_1)$ in the above inequality as
\begin{eqnarray}
c_{ij}(p_j,\mathbf{\tilde{p}}_{-j}^i) &\triangleq&
\left(1-\frac{v_j}{u_j(p_j,\mathbf{\tilde{p}}^i_{-j})}\right)\rho(y_1|\mathbf{\tilde{p}}^i)+\frac{v_j}{u_j(p_j,\mathbf{\tilde{p}}^i_{-j})}\rho(y_1|p_j,\mathbf{\tilde{p}}^i_{-j})) \\
&=& \rho(y_1|\mathbf{\tilde{p}}^i)+v_j\cdot\frac{\rho(y_1|p_j,\mathbf{\tilde{p}}^i_{-j})-\rho(y_1|\mathbf{\tilde{p}}^i)}{u_j(p_j,\mathbf{\tilde{p}}^i_{-j})} \\
&=&
\rho(y_1|\mathbf{\tilde{p}}^i)+v_j\cdot\frac{\rho(y_0|\mathbf{\tilde{p}}^i)-\rho(y_0|p_j,\mathbf{\tilde{p}}^i_{-j})}{u_j(p_j,\mathbf{\tilde{p}}^i_{-j})},
\end{eqnarray}
and define the maximum value of the coefficient $c_{ij}$ as
\begin{eqnarray}
c_{ij}^+ &\triangleq& \max_{p_j\in\mathcal{P}_j,p_j\neq \tilde{p}_j^i} c_{ij}(p_j,\mathbf{\tilde{p}}_{-j}^i) \\
&=& \rho(y_1|\mathbf{\tilde{p}}^i)+v_j\cdot\max_{p_j\in\mathcal{P}_j,p_j\neq \tilde{p}_j^i}
\frac{\rho(y_0|\mathbf{\tilde{p}}^i)-\rho(y_0|p_j,\mathbf{\tilde{p}}^i_{-j})}{u_j(p_j,\mathbf{\tilde{p}}^i_{-j})}
\end{eqnarray}

Since $\gamma_j(y_1)>\gamma_j(y_0)$, the set of inequality constraints in \eqref{eqn:IC_inactive}
\begin{eqnarray}
c_{ij}(p_j,\mathbf{\tilde{p}}_{-j}^i) \cdot \gamma_j(y_1) + (1-c_{ij}(p_j,\mathbf{\tilde{p}}_{-j}^i)) \cdot \gamma_j(y_0) \leq v_j,
\end{eqnarray}
for all $p_j>0$, is equivalent to a single constraint
\begin{eqnarray}\label{eqn:c_ij_constraint1}
c_{ij}^+ \cdot \gamma_j(y_1) + (1-c_{ij}^+) \cdot \gamma_j(y_0) \leq v_j.
\end{eqnarray}
Hence, the incentive constraints \eqref{eqn:IC_inactive} for user $j\neq i$ can be rewritten as
\begin{eqnarray}\label{eqn:IC_inactive_simplified}
\left\{\begin{array}{l} \rho(y_1|\mathbf{\tilde{p}}^i) \cdot \gamma_j(y_1) + (1-\rho(y_1|\mathbf{\tilde{p}}^i)) \cdot \gamma_j(y_0) =
\frac{v_j}{\delta} \\ c_{ij}^+ \cdot \gamma_j(y_1) + (1-c_{ij}^+) \cdot \gamma_j(y_0) \leq v_j \end{array}\right.,
\end{eqnarray}
where $\mu_j\cdot\bar{v}_j\leq \gamma_j(y) \leq \bar{v}_j,\forall y\in Y$.

The first necessary condition of $\mathcal{B}_{\mu}\subseteq\mathscr{D}(\mathcal{B}_{\mu};\delta)$ is $c_{ij}^+<0$, as stated in the following
proposition.
\begin{proposition}
If $\mathcal{B}_{\mu}\subseteq\mathscr{D}(\mathcal{B}_{\mu};\delta)$, then $c_{ij}^+<0$ for all $i\in\mathcal{N}$ and for all $j\neq i$.
\end{proposition}
\begin{IEEEproof}
If $\mathcal{B}_{\mu}\subseteq\mathscr{D}(\mathcal{B}_{\mu};\delta)$, then any payoff $\mathbf{v}$ in $\mathcal{B}_{\mu}$ should satisfy
$\mathbf{v}\in\mathscr{D}(\mathcal{B}_{\mu};\delta)$. Pick a payoff $\mathbf{\hat{v}}^i$, in which
\begin{eqnarray}
\hat{v}_j^i=\left\{\begin{array}{ll} \left(1-\sum_{k\neq i} \mu_k\right) \cdot \bar{v}_i, & j=i \\ \mu_j\cdot \bar{v}_j, & j\neq i
\end{array}\right..
\end{eqnarray}
Note that $\mathbf{\hat{v}}^i$ is the payoff profile in which every user $j\neq i$ has the smallest payoff $\mu_j\cdot \bar{v}_j$ and user $i$ has
the largest payoff $\left(1-\sum_{k\neq i} \mu_k\right) \cdot \bar{v}_i$. We show that $\mathbf{\hat{v}}^i\in\mathscr{D}(\mathcal{B}_{\mu};\delta)$
implies $c_{ij}^+<0$ for all $j\neq i$.

First, $\mathbf{\hat{v}}^i$ can only be decomposed by $\mathbf{\tilde{p}}^i$. Otherwise, suppose that $\mathbf{\hat{v}}^i$ is decomposed by
$\mathbf{\tilde{p}}^j,j\neq i$. Then the decomposition of user $i$'s payoff is
\begin{eqnarray}
\hat{v}_i^i=\delta\cdot \left(\rho(y_1|\mathbf{\tilde{p}}^j) \cdot \gamma_i(y_1) + (1-\rho(y_1|\mathbf{\tilde{p}}^j)) \cdot \gamma_i(y_0)\right).
\end{eqnarray}
Since the convex combination of $\gamma_i(y_1)$ and $\gamma_j(y_1)$ is equal to $\hat{v}_i^i/\delta$, which is strictly larger than $\hat{v}_i^i$, at
least one of $\gamma_i(y_1)$ and $\gamma_j(y_1)$ is strictly larger than $\hat{v}_i^i$. However, $\gamma_i(y)\in\mathcal{B}_{\mu}$ implies that
$\gamma_i(y)\leq \hat{v}_i^i,\forall y\in Y$, which leads to contradiction. Hence, $\mathbf{\hat{v}}^i$ can only be decomposed by
$\mathbf{\tilde{p}}^i$.

Now that $\mathbf{\hat{v}}^i$ is decomposed by $\mathbf{\tilde{p}}^i$, we focus on the incentive constraints for an arbitrary user $j\neq i$ in
\eqref{eqn:IC_inactive_simplified}. From the equality in \eqref{eqn:IC_inactive_simplified} and the requirement that $\gamma_j(y_1)>\gamma_j(y_0)$,
we have $\gamma_j(y_1)\geq \hat{v}_j^i/\delta > \hat{v}_j^i$. Then suppose that $c_{ij}^+\geq0$, in order to satisfy the inequality in
\eqref{eqn:IC_inactive_simplified}, we must have $\gamma_j(y_0)<\hat{v}_j^i$, which is contradictory to the fact that
$\gamma_j(y_0)\in\mathcal{B}_{\mu}$. Hence, we must have $c_{ij}^+<0$ for all $j\neq i$.

Since the above argument of $\mathbf{\hat{v}}^i$ applies to any $i\in\mathcal{N}$, we have $c_{ij}^+<0$ for all $i\in\mathcal{N}$ and for all $j\neq
i$.
\end{IEEEproof}

The first necessary condition that $c_{ij}^+<0$ has two implications. First, since $\rho(y_1|\mathbf{\tilde{p}}^i)$ and $v_j$ are both nonnegative,
we have
\begin{eqnarray}
\max_{p_j\in\mathcal{P}_j,p_j\neq \tilde{p}_j^i}
\frac{\rho(y_0|\mathbf{\tilde{p}}^i)-\rho(y_0|p_j,\mathbf{\tilde{p}}^i_{-j})}{u_j(p_j,\mathbf{\tilde{p}}^i_{-j})}<0,
\end{eqnarray}
where leads to Condition~1 in Theorem~\ref{theorem:CharacterizeEquilibriumPayoff} that benefit from deviation $b_{ij}<0$.

Second, to decompose $\mathbf{\hat{v}}^i$, we have
\begin{eqnarray}
c_{ij}^+ &=& \rho(y_1|\mathbf{\tilde{p}}^i)+v_j^i\cdot\max_{p_j\in\mathcal{P}_j,p_j\neq \tilde{p}_j^i}
\frac{\rho(y_0|\mathbf{\tilde{p}}^i)-\rho(y_0|p_j,\mathbf{\tilde{p}}^i_{-j})}{u_j(p_j,\mathbf{\tilde{p}}^i_{-j})} \\
&=& \rho(y_1|\mathbf{\tilde{p}}^i)+\mu_j\bar{v}_j\cdot\frac{b_{ij}}{\bar{v}_j} \\
&=& \rho(y_1|\mathbf{\tilde{p}}^i)+\mu_j\cdot b_{ij} \\
&<& 0,
\end{eqnarray}
which gives us a lower bound on $\mu_j$, namely
\begin{eqnarray}
\mu_j > \frac{\rho(y_1|\mathbf{\tilde{p}}^i)}{-b_{ij}} = \frac{1-\rho(y_0|\mathbf{\tilde{p}}^i)}{-b_{ij}}.
\end{eqnarray}
Since $\mathbf{\hat{v}}^i$ should be decomposed for all $i\in\mathcal{N}$, we have
\begin{eqnarray}
\mu_j > \max_{i\neq j} \frac{1-\rho(y_0|\mathbf{\tilde{p}}^i)}{-b_{ij}},
\end{eqnarray}
which leads to the lower bound $\underline{\mu}_j$ in Theorem~\ref{theorem:CharacterizeEquilibriumPayoff}.

\subsubsection{Incentive Constraints For The Active User}
We examine the incentive constraints for the active user $i$ in \eqref{eqn:IC_active}, which will lead to the second necessary condition (i.e.
Condition~2 in Theorem~\ref{theorem:CharacterizeEquilibriumPayoff}).

Suppose that a payoff $\mathbf{v}\in\mathcal{B}_{\mu}$ is decomposed by $\mathbf{\tilde{p}}^i$. We rewrite the incentive constraint for the active
user $i$ here
\begin{eqnarray}\label{eqn:IC_active_rewrite}
v_i &=& (1-\delta)\cdot \bar{v}_i + \delta \cdot \sum_{y\in Y} \gamma_i(y)\rho(y|\mathbf{\tilde{p}}^i) \\
    &\geq& (1-\delta)\cdot u_i(p_i,\mathbf{\tilde{p}}^i_{-i}) + \delta \cdot \sum_{y\in Y} \gamma_i(y)\rho(y|p_i,\mathbf{\tilde{p}}^i_{-i}),~\forall
    p_i\in\mathcal{P}_i. \nonumber
\end{eqnarray}
Since $\bm{\gamma}(y)\in\mathcal{B}_{\mu}$, given the inactive users' continuation payoffs $\gamma_j(y)$, the active user's continuation payoff is
determined by $\gamma_i(y)=\bar{v}_i\left(1-\sum_{j\neq i} \frac{\gamma_j(y)}{\bar{v}_j}\right)$.

First, it is not difficult to check that if $\{\gamma_j(y)\}_{j\neq i},\forall y$ satisfy the inactive users' equality constraints in
\eqref{eqn:IC_inactive_simplified}, then $\gamma_i(y)=\bar{v}_i\left(1-\sum_{j\neq i} \frac{\gamma_j(y)}{\bar{v}_j}\right)$ will satisfy the active
user's equality constraint in \eqref{eqn:IC_active_rewrite}.
\begin{eqnarray}
(1-\delta)\cdot \bar{v}_i + \delta \cdot \sum_{y\in Y} \gamma_i(y)\rho(y|\mathbf{\tilde{p}}^i)
    &=& (1-\delta)\cdot \bar{v}_i + \delta \cdot \sum_{y\in Y} \bar{v}_i\left(1-\sum_{j\neq i} \frac{\gamma_j(y)}{\bar{v}_j}\right)\rho(y|\mathbf{\tilde{p}}^i) \nonumber \\
    &=& (1-\delta)\cdot \bar{v}_i + \delta \cdot \sum_{y\in Y} \bar{v}_i\rho(y|\mathbf{\tilde{p}}^i)-\delta \cdot \sum_{y\in Y} \sum_{j\neq i} \frac{\gamma_j(y)}{\bar{v}_j}\rho(y|\mathbf{\tilde{p}}^i) \nonumber \\
    &=& \bar{v}_i-\delta \cdot \bar{v}_i \sum_{j\neq i} \sum_{y\in Y} \frac{\gamma_j(y)\rho(y|\mathbf{\tilde{p}}^i)}{\bar{v}_j} \nonumber \\
    &=& \bar{v}_i-\delta \cdot \bar{v}_i \sum_{j\neq i} \frac{v_j/\delta}{\bar{v}_j} = \bar{v}_i\left(1-\sum_{j\neq i} \frac{v_j}{\bar{v}_j}\right) =
    v_i. \nonumber
\end{eqnarray}

The inequality constraint in \eqref{eqn:IC_active_rewrite} requires that the active user $i$ has no incentive to choose another action $p_i\neq
\tilde{p}_i^i$. Although the active user $i$'s current payoff is maximized at $\mathbf{\tilde{p}}^i$, it may still have the incentive to deviate for
the following reason. Since $\gamma_j(y_1)>\gamma_j(y_0)$ for all $j\neq i$, we have $\gamma_i(y_1)<\gamma_i(y_0)$. In other words, the active user
$i$ has a larger continuation payoff when the distress signal $y_0$ is received. Hence, it may want to deviate, such that the probability of
receiving the distress signal is increased, if the increase of the expected continuation payoff outweighs the decrease of the current payoff. To
prevent the active user $i$ from deviating, we should make its continuation payoffs $\gamma_i(y_1)$ and $\gamma_i(y_0)$ as close as possible.
Equivalently, we should make the inactive users' continuation payoffs $\gamma_j(y_1)$ and $\gamma_j(y_0)$ as close as possible.

For an inactive user $j\neq i$, the closest continuation payoffs that satisfy the incentive constraints \eqref{eqn:IC_inactive_simplified} are the
ones that satisfy the inequality with equality. Hence, we can solve for the continuation payoffs as
\begin{eqnarray}\label{eqn:continuation_payoff_inactive}
\gamma_j(y_1) = \frac{\frac{1}{\delta}(1-c_{ij}^+)-(1-\rho(y_1|\mathbf{\tilde{p}}^i))}{\rho(y_1|\mathbf{\tilde{p}}^i)-c_{ij}^+}\cdot v_j,~
\gamma_j(y_0) = \frac{\rho(y_1|\mathbf{\tilde{p}}^i)-\frac{1}{\delta}c_{ij}^+}{\rho(y_1|\mathbf{\tilde{p}}^i)-c_{ij}^+}\cdot v_j.
\end{eqnarray}
Given the inactive users' continuation payoffs, we can obtain the active user's continuation payoffs $\gamma_i(y_1)$ and $\gamma_i(y_0)$. Plugging
the expression of $\gamma_j(y_1)$ and $\gamma_j(y_0)$ into the inequality in \eqref{eqn:IC_active_rewrite}, we have for all $p_i\neq \tilde{p}_i^i$,
\begin{eqnarray}\label{eqn:IC_active_rewrite}
&& v_i \geq (1-\delta)\cdot u_i(p_i,\mathbf{\tilde{p}}^i_{-i}) + \delta \cdot \sum_{y\in Y} \gamma_i(y)\rho(y|p_i,\mathbf{\tilde{p}}^i_{-i}) \nonumber \\
&\Leftrightarrow& v_i - (1-\delta)\cdot u_i(p_i,\mathbf{\tilde{p}}^i_{-i}) - \delta \cdot \sum_{y\in Y}
\bar{v}_i\left(1-\sum_{j\neq i} \frac{\gamma_j(y)}{\bar{v}_j}\right)\rho(y|p_i,\mathbf{\tilde{p}}^i_{-i}) \geq 0 \nonumber \\
&\Leftrightarrow& v_i - (1-\delta)\cdot u_i(p_i,\mathbf{\tilde{p}}^i_{-i}) - \delta\cdot\left[v_i - \bar{v}_i \cdot \sum_{j\neq i} \frac{\rho(y_1|p_i,\mathbf{\tilde{p}}^i_{-i})-c_{ij}^+}{\rho(y_1|\mathbf{\tilde{p}}^i)-c_{ij}^+}\cdot\frac{v_j}{\bar{v}_j}\cdot\left(\frac{1}{\delta}-1\right)\right] \geq 0 \nonumber \\
&\Leftrightarrow& (1-\delta)\cdot v_i - (1-\delta)\cdot u_i(p_i,\mathbf{\tilde{p}}^i_{-i}) + (1-\delta)\cdot \bar{v}_i \cdot \sum_{j\neq i} \frac{\rho(y_1|p_i,\mathbf{\tilde{p}}^i_{-i})-c_{ij}^+}{\rho(y_1|\mathbf{\tilde{p}}^i)-c_{ij}^+}\cdot\frac{v_j}{\bar{v}_j} \geq 0 \nonumber \\
&\Leftrightarrow& v_i - u_i(p_i,\mathbf{\tilde{p}}^i_{-i}) + \bar{v}_i \cdot \sum_{j\neq i} \frac{\rho(y_1|p_i,\mathbf{\tilde{p}}^i_{-i})-c_{ij}^+}{\rho(y_1|\mathbf{\tilde{p}}^i)-c_{ij}^+}\cdot\frac{v_j}{\bar{v}_j} \geq 0 \nonumber \\
&\Leftrightarrow& v_i - u_i(p_i,\mathbf{\tilde{p}}^i_{-i}) + \bar{v}_i \cdot \sum_{j\neq i} \left(1+\frac{\rho(y_1|p_i,\mathbf{\tilde{p}}^i_{-i})-\rho(y_1|\mathbf{\tilde{p}}^i)}{\rho(y_1|\mathbf{\tilde{p}}^i)-c_{ij}^+}\right)\cdot\frac{v_j}{\bar{v}_j} \geq 0 \nonumber \\
&\Leftrightarrow& \bar{v}_i\cdot\left(\frac{v_i}{\bar{v}_i}+\sum_{j\neq i} \frac{v_j}{\bar{v}_j}\right) - u_i(p_i,\mathbf{\tilde{p}}^i_{-i}) + \bar{v}_i \cdot \sum_{j\neq i} \frac{v_j/\bar{v}_j}{\rho(y_1|\mathbf{\tilde{p}}^i)-c_{ij}^+}\cdot\left(\rho(y_1|p_i,\mathbf{\tilde{p}}^i_{-i})-\rho(y_1|\mathbf{\tilde{p}}^i)\right) \geq 0 \nonumber \\
&\Leftrightarrow& \bar{v}_i - u_i(p_i,\mathbf{\tilde{p}}^i_{-i}) + \bar{v}_i \cdot \sum_{j\neq i}
\frac{\rho(y_1|p_i,\mathbf{\tilde{p}}^i_{-i})-\rho(y_1|\mathbf{\tilde{p}}^i)}{b_{ij}} \geq 0, \nonumber
\end{eqnarray}
which leads to Condition~2 in Theorem~\ref{theorem:CharacterizeEquilibriumPayoff}.

\subsubsection{Constraints On The Discount Factor}
Now we derive the necessary conditions on the discount factor. The minimum discount factor $\underline{\delta}(\bm{\mu})$ required for
$\mathcal{B}_{\bm{\mu}}$ to be a self-generating set can be solved by
\begin{eqnarray}
\underline{\delta}(\bm{\mu}) = \max_{\mathbf{v}\in\mathcal{B}_{\bm{\mu}}} \delta,~\mathrm{subject~to}~\mathbf{v} \in \mathscr{D}
(\mathcal{B}_{\bm{\mu}};\delta).
\end{eqnarray}
Since $\mathscr{D} (\mathcal{B}_{\bm{\mu}};\delta)=\cup_{i\in\mathcal{N}} \mathscr{D} (\mathcal{B}_{\bm{\mu}};\delta,\mathbf{\tilde{p}}^i)$, the
above optimization problem can be reformulated as
\begin{eqnarray}\label{eqn:delta_optimization}
\underline{\delta}(\bm{\mu}) =\max_{\mathbf{v}\in\mathcal{B}_{\bm{\mu}}} \min_{i\in\mathcal{N}} \delta,~\mathrm{subject~to}~\mathbf{v} \in
\mathscr{D} (\mathcal{B}_{\bm{\mu}};\delta,\mathbf{\tilde{p}}^i).
\end{eqnarray}

To solve the optimization problem \eqref{eqn:delta_optimization}, we explicitly express the constraint
$\mathbf{v}\in\mathscr{D}(\mathcal{B}_{\bm{\underline{\mu}}};\delta,\mathbf{\tilde{p}}^i)$ using the results derived in the previous two subsections.
The inactive users's continuation payoffs have been derived in \eqref{eqn:continuation_payoff_inactive}, which determine the active user's
continuation payoffs. Hence, the constraint $\mathbf{v}\in\mathscr{D}(\mathcal{B}_{\bm{\underline{\mu}}};\delta,\mathbf{\tilde{p}}^i)$ on discount
factor $\delta$ is equivalent to
\begin{eqnarray}
& & \bm{\gamma}(y)\in\mathcal{B}_{\mu}, \forall y\in Y,
\end{eqnarray}
which can be written explicitly as
\begin{eqnarray}
\gamma_j(y_1) &=&
\frac{\frac{1}{\delta}(1-c_{ij}^+)-(1-\rho(y_1|\mathbf{\tilde{p}}^i))}{\rho(y_1|\mathbf{\tilde{p}}^i)-c_{ij}^+}\cdot v_j \in [\mu_j\cdot\bar{v}_j,\bar{v}_j], \forall j\neq i \\
\gamma_j(y_0) &=& \frac{\rho(y_1|\mathbf{\tilde{p}}^i)-\frac{1}{\delta}c_{ij}^+}{\rho(y_1|\mathbf{\tilde{p}}^i)-c_{ij}^+}\cdot v_j  \in
[\mu_j\cdot\bar{v}_j,\bar{v}_j], \forall j\neq i \\
\gamma_i(y_1) &=&
\bar{v}_i\left(1-\sum_{j\neq i} \frac{\gamma_j(y_1)}{\bar{v}_j}\right) \in [\mu_j\cdot\bar{v}_j,\bar{v}_j] \\
\gamma_i(y_0) &=& \bar{v}_i\left(1-\sum_{j\neq i} \frac{\gamma_j(y_0)}{\bar{v}_j}\right)  \in [\mu_j\cdot\bar{v}_j,\bar{v}_j]
\end{eqnarray}

Since $\gamma_j(y_1)>\gamma_j(y_0)$, the constraints on $\gamma_j(y_1)$ and $\gamma_j(y_0)$ can be simplified as
\begin{eqnarray}\label{eqn:delta_inactive_y1}
&& \gamma_j(y_1) =
\frac{\frac{1}{\delta}(1-c_{ij}^+)-(1-\rho(y_1|\mathbf{\tilde{p}}^i))}{\rho(y_1|\mathbf{\tilde{p}}^i)-c_{ij}^+}\cdot v_j \leq \bar{v}_j \\
&\Leftrightarrow& \delta\geq \frac{1-c_{ij}^+}{1-c_{ij}^+ + \left(\frac{\bar{v}_j}{v_j}-1\right)(\rho(y_1|\mathbf{\tilde{p}}^i)-c_{ij}^+)},
\end{eqnarray}
and
\begin{eqnarray}\label{eqn:delta_inactive_y0}
&& \gamma_j(y_0) = \frac{\frac{1}{\delta}(1-c_{ij}^+)-(1-\rho(y_1|\mathbf{\tilde{p}}^i))}{\rho(y_1|\mathbf{\tilde{p}}^i)-c_{ij}^+}\cdot v_j \geq
\mu_j\cdot\bar{v}_j.
\end{eqnarray}
Note that the constraint \eqref{eqn:delta_inactive_y0} will be satisfied as long as $c_{ij}^+<0$.

Since $\gamma_i(y_1)<\gamma_i(y_0)$, the constraints on $\gamma_i(y_1)$ and $\gamma_i(y_0)$ can be simplified as
\begin{eqnarray}\label{eqn:delta_active_y1}
\gamma_i(y_1) \geq \mu_i\cdot \bar{v}_i \Leftrightarrow \delta\geq \frac{1}{1+\frac{v_i}{\bar{v}_i}\frac{1-\mu_i}{\sum_{j\neq i}
\frac{1-c_{ij}^+}{\rho(y_1|\mathbf{\tilde{p}}^i)-c_{ij}^+}\cdot\frac{v_j}{\bar{v}_j}}},
\end{eqnarray}
and
\begin{eqnarray}
\gamma_i(y_0) \leq \bar{v}_i.
\end{eqnarray}
Note that the above constraint on $\gamma_i(y_0)$ is satisfied as long as \eqref{eqn:delta_inactive_y0} is satisfied for all $j\neq i$. Note also
that the constraint \eqref{eqn:delta_inactive_y1} is satisfied as long as \eqref{eqn:delta_active_y1} is satisfied.

To sum up, the discount factor needs to satisfy the following constraint:
\begin{eqnarray}
\delta\geq \frac{1}{1+\frac{v_i}{\bar{v}_i}\frac{1-\mu_i}{\sum_{j\neq i}
\frac{1-c_{ij}^+}{\rho(y_1|\mathbf{\tilde{p}}^i)-c_{ij}^+}\cdot\frac{v_j}{\bar{v}_j}}}=\frac{1}{1+\frac{v_i}{\bar{v}_i}\frac{1-\mu_i}{\sum_{j\neq i}
\frac{1-c_{ij}^+}{-b_{ij}}}}=\frac{1}{1+\frac{v_i}{\bar{v}_i}\frac{1-\mu_i}{\sum_{j\neq i}
\left(\frac{\rho(y_0|\mathbf{\tilde{p}}^i)}{-b_{ij}}+\frac{v_j}{\bar{v}_j}\right)}}.
\end{eqnarray}

Hence, the optimization problem \eqref{eqn:delta_optimization} is equivalent to
\begin{eqnarray}\label{eqn:delta_matrix}
\underline{\delta}(\bm{\mu}) = \max_{\mathbf{v}\in\mathcal{B}_{\bm{\mu}}} \min_{i\in\mathcal{N}} x_i(\mathbf{v}),
\end{eqnarray}
where
$$
x_{i}(\mathbf{v})\triangleq \frac{1}{1+\frac{v_i}{\bar{v}_i}\frac{1-\mu_i}{\sum_{j\neq i}
\left(\frac{\rho(y_0|\mathbf{\tilde{p}}^i)}{-b_{ij}}+\frac{v_j}{\bar{v}_j}\right)}}.
$$
Since $x_i(\mathbf{v})$ is decreasing in $v_i$ and increasing in $v_j,\forall j\neq i$, the payoff $\mathbf{v}^*$ that maximizes
$\min_{i\in\mathcal{N}} x_i(\mathbf{v})$ must satisfy $x_i(\mathbf{v}^*)=x_j(\mathbf{v}^*)$ for all $i$ and $j$. Now we find the payoff
$\mathbf{v}^*$ such that $x_i(\mathbf{v}^*)=x_j(\mathbf{v}^*)$ for all $i$ and $j$.

Define $z\triangleq \frac{v_i}{\bar{v}_i}\frac{1-\mu_i}{\sum_{j\neq i}
\left(\frac{\rho(y_0|\mathbf{\tilde{p}}^i)}{-b_{ij}}+\frac{v_j}{\bar{v}_j}\right)}=\frac{v_i}{\bar{v}_i}\frac{1-\mu_i}{1-\frac{v_i}{\bar{v}_i}+\sum_{j\neq
i} \frac{\rho(y_0|\mathbf{\tilde{p}}^i)}{-b_{ij}}},\forall i\in\mathcal{N}$. Then we can solve for $\frac{v_i}{\bar{v}_i}$ as follows
\begin{eqnarray}
\frac{v_i}{\bar{v}_i} = \frac{z\left(1+\sum_{j\neq i} \frac{\rho(y_0|\mathbf{\tilde{p}}^i)}{-b_{ij}}\right)+\mu_i}{1+z}.
\end{eqnarray}
Since $\sum_{i\in\mathcal{N}} \frac{v_i}{\bar{v}_i}=1$, we can solve for $z$ as
\begin{eqnarray}
z = \frac{1-\sum_{i\in\mathcal{N}} \mu_i}{N-1+\sum_{i\in\mathcal{N}} \sum_{j\neq i} \frac{\rho(y_0|\mathbf{\tilde{p}}^i)}{-b_{ij}}}.
\end{eqnarray}
Hence, the minimum discount factor is $\underline{\delta}(\bm{\mu}) = \frac{1}{1+z}$, which leads to Condition~3 in
Theorem~\ref{theorem:CharacterizeEquilibriumPayoff}.

\subsection{Necessary Conditions Are Also Sufficient}
In the previous subsection, we have derived three necessary conditions for the set $\mathcal{B}_{\bm{\mu}}$ to be self-generating. Now we show that
the three necessary conditions are also sufficient for $\mathcal{B}_{\bm{\mu}}$ to be self-generating.

Given any payoff $\mathbf{v}\in\mathcal{B}_{\bm{\mu}}$, we can determine the action profile $\mathbf{\tilde{p}}^i$ that decomposes it and the
corresponding continuation payoffs based on the results in the previous subsection. First, the action profile $\mathbf{\tilde{p}}^i$ that decomposes
$\mathbf{v}$ is determined by
\begin{eqnarray}
i=\arg\min_{j\in\mathcal{N}} x_j(\mathbf{v}) = \arg \max_{j\in\mathcal{N}} \frac{v_j}{\bar{v}_j}\frac{1-\mu_j}{1-\frac{v_j}{\bar{v}_j}+\sum_{k\neq j}
\frac{\rho(y_0|\mathbf{\tilde{p}}^j)}{-b_{jk}}}.
\end{eqnarray}
Then we determine the continuation payoffs as
\begin{eqnarray}\label{eqn:continuation_payoff_final}
\left\{\begin{array}{l}\gamma_j(y_1) =
\frac{\frac{1}{\delta}(1-c_{ij}^+)-(1-\rho(y_1|\mathbf{\tilde{p}}^i))}{\rho(y_1|\mathbf{\tilde{p}}^i)-c_{ij}^+}\cdot v_j \leq \bar{v}_j, \forall j\neq i, \\
\gamma_j(y_0) = \frac{\frac{1}{\delta}(1-c_{ij}^+)-(1-\rho(y_1|\mathbf{\tilde{p}}^i))}{\rho(y_1|\mathbf{\tilde{p}}^i)-c_{ij}^+}\cdot v_j \geq
\mu_j\cdot\bar{v}_j, \forall j\neq i, \\
\gamma_i(y)=\bar{v}_i\left(1-\sum_{j\neq i} \frac{\gamma_j(y)}{\bar{v}_j}\right), \forall y\in Y
\end{array}\right..
\end{eqnarray}

Conditions 1 and 2 ensure that the incentive constraints for the active user \eqref{eqn:IC_active} and the inactive users \eqref{eqn:IC_inactive} are
satisfied by setting the continuation payoffs as above. Condition~3 on the discount factor $\delta$ ensures that the above continuation payoff
$\bm{\gamma}(y)\in\mathcal{B}_{\bm{\mu}}$. Hence, any payoff $\mathbf{v}\in\mathcal{B}_{\bm{\mu}}$ is decomposable on set $\mathcal{B}_{\bm{\mu}}$
with respect to discount factor $\delta\geq\underline{\delta}(\bm{\mu})$. Then $\mathcal{B}_{\bm{\mu}}$ is self-generating, and any payoff in
$\mathcal{B}_{\bm{\mu}}$ is an equilibrium payoff.

\section{Proof of Theorem~\ref{theorem:EquilibriumStrategy}}\label{proof:EquilibriumStrategy}
We have characterized the largest set of Pareto optimal equilibrium payoffs $\mathcal{B}_{\bm{\underline{\mu}}}$. In the algorithm in Table~II, we
start with the target payoff $\mathbf{v}^\star\in\mathcal{B}_{\bm{\underline{\mu}}}$ as the average payoff at period $0$, and decompose it into a
current payoff and a continuation payoff. The decomposition tells us what action profile to play in period $0$. Then we decompose the continuation
payoff and determine the action profile to play in period $1$. By performing the decomposition in every period, we can determine what action profile
to play given any signal at every period.

Specifically, suppose that the continuation payoff at period $t$ is $\mathbf{v}(t)$. Then the action profile $\mathbf{\tilde{p}}^i$ to decompose
$\mathbf{v}(t)$ is determined by
\begin{eqnarray}
i^*=\arg\min_{j\in\mathcal{N}} x_j(\mathbf{v}(t)) = \arg \max_{j\in\mathcal{N}}
\frac{v_j(t)}{\bar{v}_j}\frac{1-\mu_j}{1-\frac{v_j(t)}{\bar{v}_j}+\sum_{k\neq j} \frac{\rho(y_0|\mathbf{\tilde{p}}^j)}{-b_{jk}}},
\end{eqnarray}
where $\frac{v_j(t)}{\bar{v}_j}\frac{1-\mu_j}{1-\frac{v_j(t)}{\bar{v}_j}+\sum_{k\neq j} \frac{\rho(y_0|\mathbf{\tilde{p}}^j)}{-b_{jk}}}$ is exactly
user $j$'s index $\alpha_j(t)$. Then we can determine the continuation payoff $\mathbf{v}(t+1)$ according to \eqref{eqn:continuation_payoff_final}.

%
%

\end{document}